\documentclass[%
 reprint,footinbib,
superscriptaddress,
 floatfix,
 amsmath,amssymb,
 aps,
prapp, 
longbibliography,
]{revtex4-2}

\usepackage{graphicx}  
\usepackage{dcolumn}   
\usepackage{bm}        
\usepackage{amssymb}   
\usepackage{amsmath}
\usepackage{color}
\usepackage[colorlinks=true, pdfstartview=FitV, linkcolor=blue, citecolor=blue, urlcolor=blue]{hyperref} 
\usepackage{epstopdf}
\usepackage{siunitx} 
\usepackage{tabularx} 
\usepackage{xspace}
\usepackage{braket}
\usepackage{color}
\usepackage{setspace}
\usepackage{mdframed}
\usepackage{framed}
\usepackage{multirow}
\usepackage{setspace}
\usepackage[export]{adjustbox}
\usepackage{varwidth}
\usepackage{float}

\usepackage{lipsum}

\newcommand{\affilETH}{Laboratory for Solid State Physics, ETH Z\"{u}rich, 8093 Z\"urich, Switzerland.}
\newcommand{\affilNBI}{Niels Bohr Institute, University of Copenhagen, 2100 Copenhagen, Denmark.}
\newcommand{\affilNBIHYQ}{Center for Hybrid Quantum Networks, Niels Bohr Institute, University of Copenhagen, 2100 Copenhagen, Denmark.}

\newcommand{\affilKonstanz}{Department of Physics, University of Konstanz, D-78457 Konstanz, Germany.}
\newcommand{\affilETHQuantum}{Quantum Center, ETH Zurich, 8093 Zurich, Switzerland.}
\newcommand{\ITP}{Institute for Theoretical Physics, ETH Z{\"u}rich, 8093 Z{\"u}rich, Switzerland.}

\begin{document}


\title{Strong parametric coupling between two ultra-coherent membrane modes}

\author{David H\"alg}
\affiliation{\affilETH}
\author{Thomas Gisler}
\affiliation{\affilETH}
\author{Eric C. Langman}
\affiliation{\affilNBI}
\affiliation{\affilNBIHYQ}
\author{Shobhna Misra}
\affiliation{\affilETH}
\author{Oded Zilberberg}
\affiliation{\ITP}
\affiliation{\affilKonstanz}
\author{Albert Schliesser}
\affiliation{\affilNBI}
\affiliation{\affilNBIHYQ}
\author{Christian L. Degen}
\affiliation{\affilETH}
\affiliation{\affilETHQuantum}
\author{Alexander Eichler}
\email[Corresponding author: ]{eichlera@ethz.ch}
\affiliation{\affilETH}



\date{\today} 

\begin{abstract}
We demonstrate parametric coupling between two modes of a silicon nitride membrane. We achieve the coupling by applying an oscillating voltage to a sharp metal tip that approaches the membrane surface to within a few \SI{100}{\nano\meter}. When the voltage oscillation frequency is equal to the mode frequency difference, the modes exchange energy periodically and much faster than their free energy decay rate. This flexible method can potentially be useful for rapid state control and transfer between modes, and is an important step towards parametric spin sensing experiments with membrane resonators.
\end{abstract}

\maketitle


Optomechanical resonators made from silicon nitride are a key resource for future technologies~\cite{Verbridge_2008,Anetsberger_2010,Reetz_2019,Reinhardt_2016,Tsaturyan_2017,Ghadimi_2018,Beccari_2021_hierarchical,Beccari_2021}. Owing to their very large quality factors, these devices are highly coherent and extremely sensitive to external forces~\cite{Halg_2021}. They can interact with mechanical, electrical, optical and magnetic signals, and their mechanical mode shapes and resonance frequencies can be tailored to specific situations. Promising applications of such resonators include demonstrations of fundamental light-matter interactions~\cite{Peterson_2016,Rossi_2018,mason_continuous_2019}, gravitational-wave detection~\cite{Page_2021_gravitational}, quantum-coherent signal conversion between the microwave and optical domains~\cite{Bagci_2014, Andrews_2014}, scanning force microscopy~\cite{Halg_2021}, and nanoscale magnetic resonance imaging~\cite{Fischer_2019,Kosata_2020}.

One crucial ingredient for harnessing optomechanical resonators for applications involves parametric coupling, also known as parametric frequency conversion, mode locking or three-wave mixing. This method for coupling non-degenerate modes, employed since a long time in electronics and nonlinear optics~\cite{Boyd_2020}, can also be applied to nanomechanical resonators~\cite{Dougherty_1996}. In parametric coupling, two modes with frequencies $f_i$ and $f_j$ are coupled through a pump tone at the frequency difference $f_p = \vert f_j - f_i\vert$. Due to the pump tone, the modes experience each other's oscillations as resonant forces, similar to coupled degenerate resonators. However, implementing efficient parametric coupling with ultra-coherent optomechanical devices is difficult -- the very shielding from the environment that facilitates exceptionally high quality factors~\cite{Tsaturyan_2017,Ghadimi_2018} makes the implementation of a strong pump tone challenging. As a result, strong parametric coupling~\cite{Faust_2013,Okamoto_2013}, where the rate of energy exchange between two nondegenerate modes exceeds their individual decay rates, has so far not been demonstrated with ultra-high $Q$ resonators.

In this work, we demonstrate strong parametric coupling between two shielded mechanical modes in a silicon nitride membrane. We circumvent the shielding issue with a sharp metallic tip that approaches the membrane to a few \SI{100}{\nano\meter}, enabling local and precise control of the electrical interaction force gradient. With this control, we demonstrate parametric coupling with a time $t_\Delta = \SI{4.8}{\second}$ required to transmit energy from one mode to the other and back, roughly 7 times faster than the mechanical decay time $\tau$. Our method can be further improved by increasing the power delivered by the electrical pump tone, the electrical make-up of the membrane surface, and the quality factor of the mechanical modes, leading ultimately to quantum coherent state transfer between non-degenerate modes.

        
        
        \begin{figure*}
        \includegraphics[width=\textwidth]{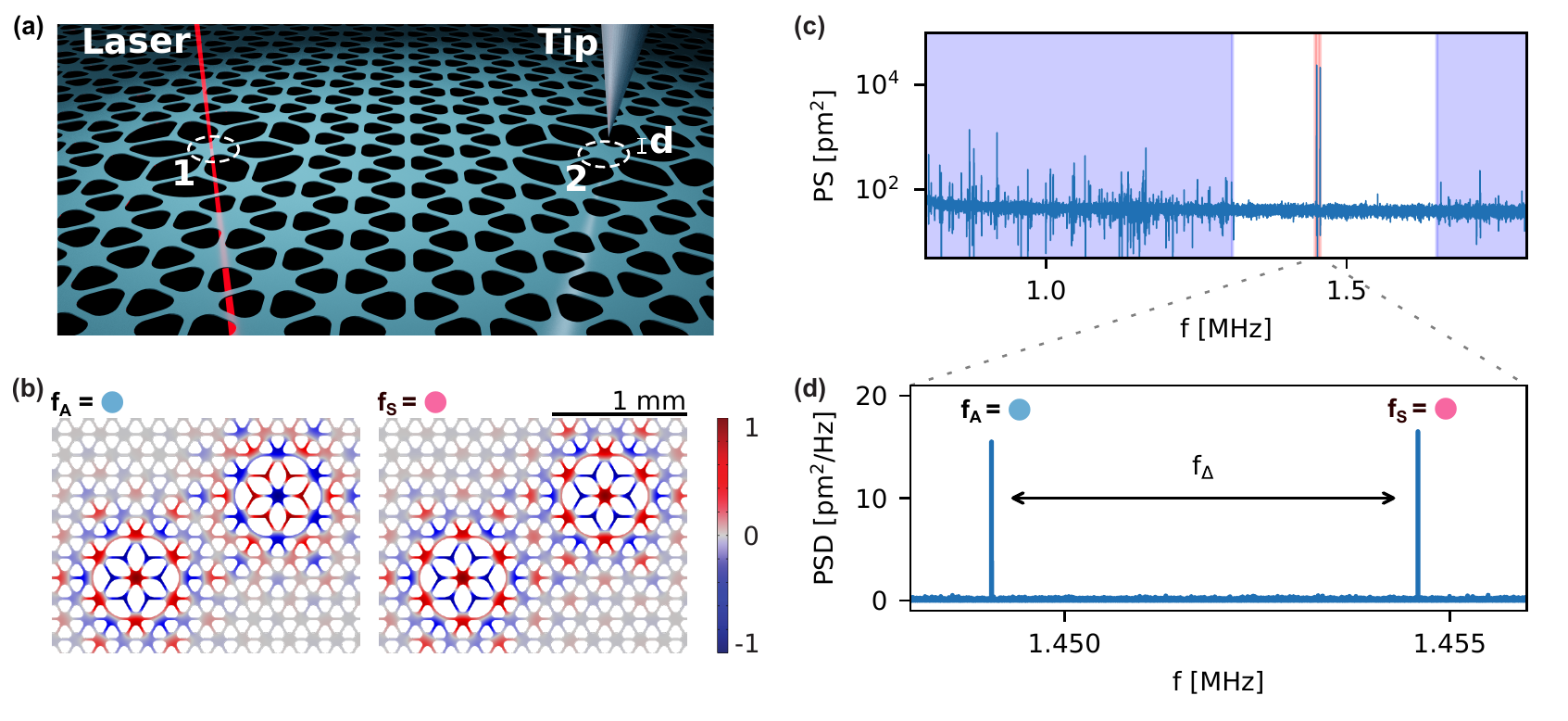} 
        \caption{\textbf{Setup and Characterization.}
        (a)~Sketch of the setup, showing the phononic crystal hole pattern with two defect sites labeled 1 and 2. The interferometric readout and driving lasers are centered on defect site 1 and the scanning tip is centered at a distance $d$ above defect 2. (b)~Simulated displacement of the anti-symmetric (left) and symmetric (right) membrane modes in arbitrary units. (c)~Displacement power spectrum obtained by slowly sweeping the frequency of the laser drive. Shaded blue region: modes outside of the bandgap. Shaded red region: frequency range shown in (d). (d)~Displacement power spectral density of the membrane modes shown in (b), driven by thermomechanical force noise. The left and right peaks are the anti-symmetric and symmetric modes, respectively, color coded in light blue and pink throughout the paper.
        }
        \label{fig:Setup}
        \end{figure*}


At the heart of our experiment is a silicon nitride membrane resonator whose displacement is detected with a power-balanced laser interferometer at \SI{1550}{\nano\meter}, see Fig.~\ref{fig:Setup}(a). The time-dependent light intensity is transformed into an electrical signal and measured with a \textsc{Zurich Instruments HF2LI} lock-in amplifier. A sharp (rigid) metal tip can be scanned over the membrane surface with a three-axis stage~\cite{Halg_2021}. The \SI{20}{\nano\meter}-thick membrane features a hole pattern that implements a phononic crystal~\cite{Tsaturyan_2017}. Two `lotus' defect \cite{Seis_2021} islands in the pattern with a separation of roughly \SI{1.2}{\milli\meter} and with quality factors $Q = \SI{1.4e8}{}$ act as isolated resonators 1 and 2. The symmetric (S) and antisymmetric (A) normal modes of these resonators, shown in Fig.~\ref{fig:Setup}(b), extend over both defect sites.\cite{Catalini_2020} In the spectra in Fig.~\ref{fig:Setup}(c) and (d), the normal modes appear as narrow lines at frequencies $f_{A,S}$ inside the bandgap created by the phononic crystal.

        \begin{figure}
        \includegraphics[width=1.02\columnwidth]{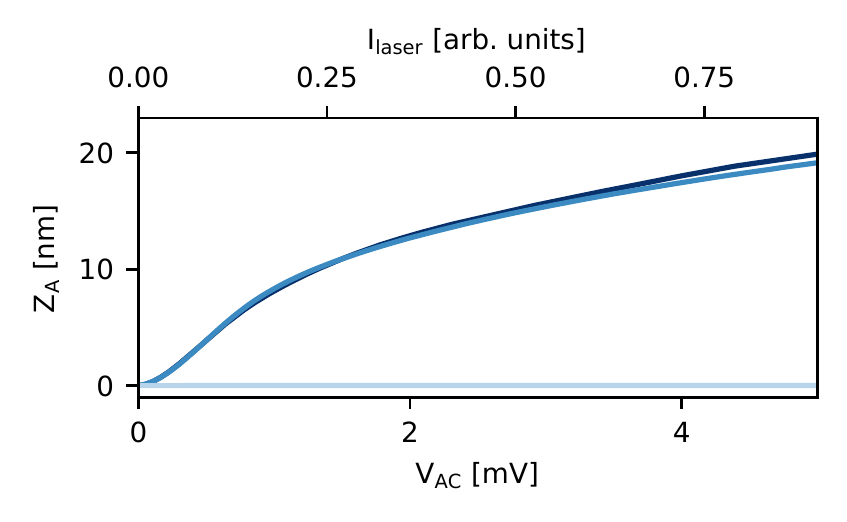} 
        \caption{\textbf{Comparison of different driving methods.}
        The amplitude $Z_A$ of the anti-symmetric mode driven with a photon pressure modulated at $f_d=f_A$ (dark blue), with a voltage applied to the tip at $f_d=f_A$ (medium blue) and with a tip voltage at $f_d=\frac{1}{2}f_A$ (light blue).
        }
        \label{fig:drives}
        \end{figure}

We can apply resonant driving forces to the two modes and measure their oscillatory response, see Fig.~\ref{fig:drives}. First, we drive mode A with the photon pressure of a \SI{980}{\nano\meter} laser whose intensity $I_\mathrm{laser}$ is modulated at $f_d = f_A$~\cite{Halg_2021}, generating a force $F_A\propto I_\mathrm{laser}$. Second, we drive the same mode by modulating the voltage applied to the scanning tip as $V_\mathrm{tip} = V_{\mathrm{AC}}\cos(2\pi f_d t)$~\cite{Halg_2021}. The response for $f_d = f_A$ exhibits the same nonlinear saturation at large amplitudes as obtained with the photon drive. There are three possibilities how an electrical force can arise: (i)~for a conducting sample, a Coulomb force is generated between the charged tip and mirror charges on the membrane surface. Since our membrane is made from a dielectric, we think this to be an unlikely option. (ii)~The dielectric membrane material can become polarized in response to the applied voltage~\cite{Unterreithmeier2009}. For both (i) and (ii), we expect $F_A \propto V_{\mathrm{AC}}^2$ and that the force appears at $2f_d$ instead of $f_d$. However, our measurements indicate $F_A\propto V_{\mathrm{AC}}$ and the vibrational response when driving at $f_d = f_A/2$, shown in Fig.~\ref{fig:drives} for comparison, is negligible.
(iii)~We finally propose that the force is caused by the interaction of the tip with `voltage patches' on the membrane surface~\cite{camp_macroscopic_1991, burnham_work-function_1992, rossi_observations_1992, speake_forces_2003, gaillard_method_2006, robertson_kelvin_2006, zuniga_polarity_2005}. A similar interaction was recently observed with a different nanomechanical system~\cite{heritier_2021}. This model can qualitatively explain all phenomena in Fig.~\ref{fig:drives}. Additional data in the Supplemental material~\cite{Supplement} support this interpretation. A quantitative analysis will be the subject of future work.

Our system can be modeled with a set of coupled equations of the form
\begin{subequations}
\label{eq:EOM}
\begin{align}
    \ddot{z}_{A} + \omega_{A}^2 z_{A} + \Gamma_{A}\dot{z}_{A} - J z_{S} = F_{A}/m_{A} \label{eq:EOM1}\\
    \ddot{z}_{S} + \omega_{S}^2 z_{S} + \Gamma_{S}\dot{z}_{S} - J z_{A} = F_{S}/m_{S}\,.\label{eq:EOM2}
\end{align}
\end{subequations}
Here $z_i$ is the displacement (with dots signifying time derivatives), $\omega_i = 2\pi f_i$ the angular resonance frequency, $\Gamma = \frac{\omega_i}{Q_i}$ the dissipation coefficient of mode $i$ with quality factor $Q_i$, $m_i$ the effective mass, and $F_i$ the force acting on mode $i$. For perfectly degenerate defect resonators 1 and 2, the normal modes S and A are fully decoupled. Finite detuning between the defect resonators, in contrast, leads to finite coupling $J$~\cite{Frimmer_2014}. Consequently, a modulation of the detuning between the resonators at a frequency $f_p$ chosen to be equal to $f_{\Delta} = f_S-f_A$ or $f_{\Sigma} = f_S+f_A$ leads to a coupling term $J(t) z_j = g \cos(2\pi f_p t) z_j$ that has a component which is resonant with $f_i$~\cite{Frimmer_2014, Kosata_2020}. This `parametric coupling' can be used for an efficient energy exchange between modes at very different frequencies~\cite{Dougherty_1996,Faust_2013,Okamoto_2013_NP} and represents a bosonic, classical analogue to Rabi oscillations.

    \begin{figure}
    \includegraphics[width=1.02\columnwidth]{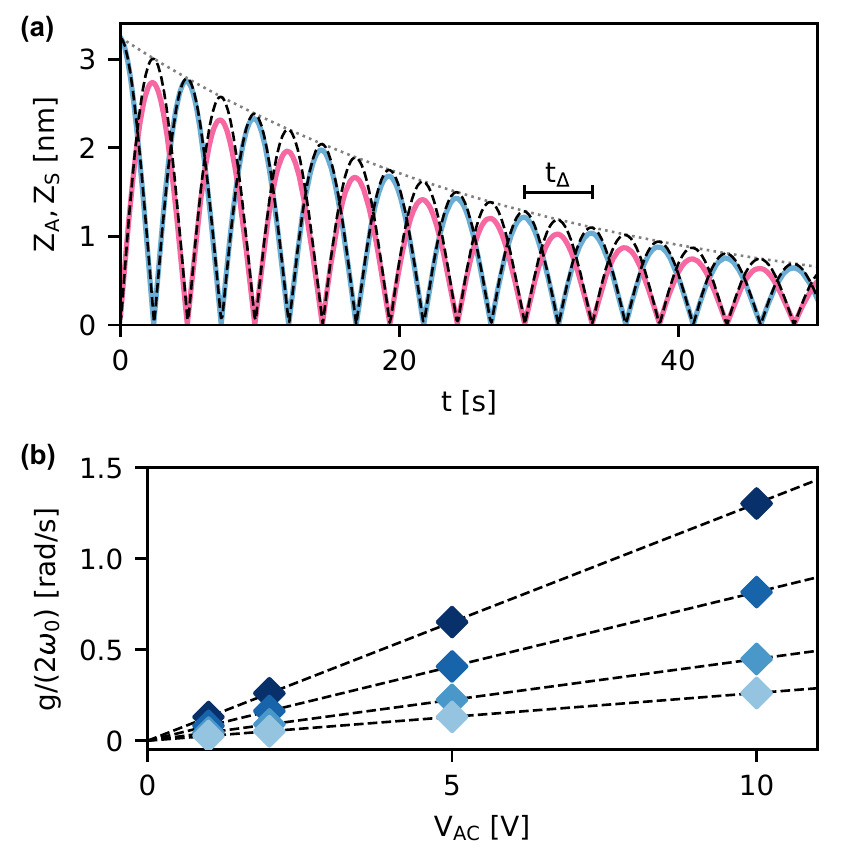} 
    \caption{\textbf{Parametric pumping at $f_\Delta$.}
    (a)~Parametric pumping with $V_\mathrm{AC}=\SI{10}{\volt}$ at $d=\SI{100}{\nano\meter}$ and with initial conditions $Z_S(0) = 0$, $Z_A(0) = \SI{3.25}{\nano\meter}$. Blue and pink: measured amplitudes $Z_i = \sqrt{u_i^2+v_i^2}$ of the anti-symmetric and symmetric mode, respectively. Black dashed line: theory fit using Eq.~\eqref{eq:slow_flow} with $g/(2\omega_0) = \SI{1.3}{\radian\per\second}$ and $\delta_{S,A} = 0$. Grey dotted line: exponential decay with the average of the ringdown times $\tau_i = Q_i/\pi f_i$. (b)~Coupling strength $g/(2\omega_0) $ versus tip voltage. From light to dark blue at each voltage: $d = \SI{1000}{\nano\meter}$, \SI{500}{\nano\meter}, \SI{200}{\nano\meter}, and \SI{100}{\nano\meter}. Dashed lines are linear fits.
    }
    \label{fig:parametricRD}
    \end{figure}

We derive a formal description of the parametric coupling process via the averaging method~\cite{Supplement}. In a frame rotating at $f_i^\mathrm{av}\approx f_i$, the displacement $z_i$ can be expressed in terms of in-phase and out-of-phase oscillation amplitudes $u_i$ and $v_i$, respectively. The time evolution of the system is then obtained from the coupled slow-flow equations
\begin{align}\label{eq:slow_flow}
\begin{bmatrix}
\dot{u}_S \\
\dot{v}_S \\
\dot{u}_A \\
\dot{v}_A 
\end{bmatrix}
= \frac{1}{2}
\begin{bmatrix}
-\Gamma & 2\delta_S & 0 & \frac{\sigma g}{2\omega_0} \\
-2\delta_S & -\Gamma & -\frac{g}{2\omega_0} & 0 \\
0 & \frac{\sigma g}{2\omega_0} & -\Gamma & 2\delta_A \\
-\frac{g}{2\omega_0} & 0 & -2\delta_A & -\Gamma
\end{bmatrix}
\begin{bmatrix}
u_S \\
v_S \\
u_A \\
v_A 
\end{bmatrix}
+
\begin{bmatrix}
0 \\
\frac{-F_S}{2m\omega_0}\\
0 \\
\frac{-F_A}{2m\omega_0}
\end{bmatrix},
\end{align}
where $\delta_i/2\pi = f_i^\mathrm{av} - f_i$ are small detunings from the ideal rotating frame, $\omega_0=(\omega_S+\omega_A)/2$, and we use the notation $\sigma = 1$ ($\sigma = -1$) for driving at $f_p = f_\Delta$ ($f_p = f_\Sigma$). Our choice of zero phase offset for $F_A$ and $F_S$ leads to force terms that act solely on the $v_A$ and $v_S$ coordinates on resonance. The result we obtain for pumping at $f_p = f_\Delta$ is identical to the one previously derived with a different method~\cite{Frimmer_2014}. Below, we concentrate on results obtained by driving at $f_\Delta$. We refer the reader to the supplement for an example of driving at $f_\Sigma$~\cite{Supplement}.

A time-dependent detuning $J$ between the resonators 1 and 2 is achieved by applying an electrical drive tone $V_\mathrm{tip} = V_{\mathrm{AC}}\cos(2\pi f_p t)$ to resonator 2 via the scanning tip. The corresponding force gradient $-\frac{\delta F_\mathrm{el}(t)}{\delta z}$ induces an electrical spring constant $k_\mathrm{el}(t)$ that modifies the resonator's resonance frequency $f_2$ as a function of time. As the microscopic origin of the electrical force $F_\mathrm{el}$ is still under scrutiny (see above), we cannot calculate the conversion between $V_\mathrm{tip}$ and $F_\mathrm{el}$ from first principles. Instead, we will, in the following, extract the parametric modulation depth $g$ from measurements.


In Fig.~\ref{fig:parametricRD}, we experimentally investigate parametric coupling at $f_p=f_\Delta$ in the absence of an external force. In Fig.~\ref{fig:parametricRD}(a), the anti-symmetric mode is initially driven to high amplitude by a resonant laser drive. The laser drive is then switched off and a parametric pump tone at $f_\Delta$ is switched on. The two modes exchange energy periodically in additional to an overall ringdown.
The theoretical fit yields a coupling frequency of $\frac{g}{2\omega_0} = \frac{2\pi}{t_\Delta}=\SI{1.3}{\radian\per\second}$, which is the highest coupling frequency we achieved with this device (for $V_{\text{AC}}= \SI{10}{\volt}$ and $d=\SI{100}{\nano\meter}$).
In Fig.~\ref{fig:parametricRD}(b), we collect the measured values of $\frac{g}{2\omega_0}$ for different $V_\mathrm{AC}$ and $d$. We find a linear dependence on $V_\mathrm{AC}$, while the increase in coupling frequency is stronger than linear with $d$.

    \begin{figure}
    \includegraphics[width=1.02\columnwidth]{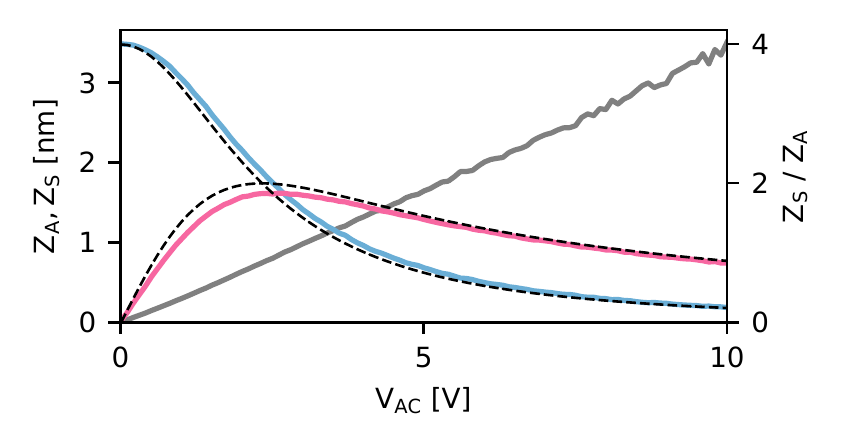} 
    \caption{\textbf{Steady-state amplitudes.}
    Mode amplitudes as a function of $V_\mathrm{AC}$ with a constant photon pressure drive at $f_d = f_A$ and parametric pumping at $f_\Delta$ at a distance $d=$\SI{1000}{\nano\meter} (left axis). Blue and pink: measured amplitude of the anti-symmetric and symmetric mode, respectively. Black dashed lines: solutions of Eq.~\eqref{eq:slow_flow} for $\dot{u}_{S,A} = \dot{v}_{S,A} = 0$. Grey: $Z_S/Z_A$ extracted from the measurements (right axis).
    }
    \label{fig:SteadyState}
    \end{figure}

When applying a resonant laser (photon pressure) drive to mode A and a parametric pump tone at $f_\Delta$ simultaneously, the system reaches an out-of-equilibrium steady state after a time $t \gg Q/\pi f$~\cite{Dougherty_1996,Kosata_2020}. The amplitudes of this steady state depend on $V_\mathrm{AC}\propto g$, offering an alternative way to measure the coupling. The measured and calculated results are shown in Fig.~\ref{fig:SteadyState}. For increasing parametric coupling, the steady-state amplitudes begin to equalize and cross at $\approx\SI{2.5}{\volt}$, whereafter mode A has the \textit{smaller} amplitude of the two modes. To understand this, we need to remember that the phase of mode S follows the force exerted by the coupling term with a phase lag of $-\pi/2$, and that mode S in turn drives mode A with a second phase lag of $-\pi/2$. In total, mode A damps itself through this feedback loop, and its amplitude follows the simple closed form (see SI)
\begin{align}\label{eq:A_steady}
Z_A = \sqrt{u_A^2 + v_A^2} = -i \frac{F_A}{m\omega_0\Gamma} \left[1 + \left(\frac{g}{2\omega_0\Gamma}\right)^2\right]^{-1}
\end{align}
while the ratio between the two amplitudes is
\begin{align}\label{eq:ratio_steady}
\frac{Z_S}{Z_A} = -i \frac{g}{2\omega_0\Gamma}\,.
\end{align}
We find excellent correspondence between the measurements and Eqs.~\eqref{eq:A_steady} and \eqref{eq:ratio_steady}, using $F_A$ and $g$ as free parameters. The value $g/2\omega_0 = \SI{0.25}{\radian\per\second}$ obtained for $V_{AC} = \SI{10}{\volt}$ and $d=\SI{1000}{\nano\meter}$ is the same (within fit errors) as the one we found for identical conditions in Fig.~\ref{fig:parametricRD}.

The agreement between experiment and theory serves as a foundation for future applications. In particular, the linear relationship between the two mode amplitudes in Fig.~\ref{fig:SteadyState} is an important `dress rehearsal' for nuclear spin detection with a membrane resonator~\cite{Kosata_2020}. In this proposal, the parametric pumping is provided by an ensemble of periodically inverted nuclear spins, whose number can then be estimated from the amplitude of the undriven mode ($Z_S$ in Fig.~\ref{fig:SteadyState}). Our experiment shows that the parametric energy exchange between modes follows a simple mathematical form and is not significantly affected by the modes' nonlinearities~\cite{Catalini_2020,Kosata_2020} in the studied amplitude range.

Parametric mode conversion can also be of potential use for purely electrical applications of scanning force microscopy. For instance, a deeper understanding of the microscopic origin of mechanical damping will require highly resolved and ultrasensitive imaging of trapped charges on resonator surfaces~\cite{camp_macroscopic_1991, burnham_work-function_1992, rossi_observations_1992, speake_forces_2003, gaillard_method_2006, robertson_kelvin_2006, zuniga_polarity_2005,heritier_2021}. For a single charge, the force sensitivity of the parametric frequency conversion method (cf. Fig.~\ref{fig:SteadyState}) can be compared to the direct-drive method. Direct driving between a single surface charge $q_\mathrm{surf}$ and a spherical tip apex with a charge $q_\mathrm{tip}$ results in a Coulomb force
\begin{align}
    F_S = \frac{1}{4\pi \epsilon_0}\frac{q_\mathrm{tip} q_\mathrm{surf}}{(d+R)^2}\,,
\end{align}
where $\epsilon_0$ is the permittivity of free space and $d+R$ the separation between the effective charge positions, with $R$ the tip apex radius. The force is the first derivative of the resonator's electric potential $E_\mathrm{c}$ in the presence of a charged tip. In contrast, the parametric method probes the second derivative of $E_\mathrm{c}$ and yields
\begin{align}\label{eq:parametric_electrical}
    F_\mathrm{par} = J z_A = -\frac{1}{2\pi \epsilon_0}\frac{q_\mathrm{tip} q_\mathrm{surf}}{(d+R)^3} z_A\,.
\end{align} 
The two forces become equal for $z_A = (d+R)/2$, which can in principle be attained for $d>R$. The calculation generally shows that the two methods do not differ by much in sensitivity for $z_A \to d$. Inserting into Eq.~\eqref{eq:parametric_electrical} the numerical values $q_\mathrm{tip} = q_\mathrm{surf} = \SI{0.16}{\atto\coulomb}$ (one elementary charge), $d = R = \SI{50}{\nano\meter}$ and $z_S = \SI{10}{\nano\meter}$, we obtain $F_\mathrm{par} = \SI{4.6}{\femto\newton}$. This force should be routinely detected with our ultrasensitive membrane resonators whose force noise will drop below \SI{1}{\atto\newton\hertz^{-1/2}} at cryogenic temperatures~\cite{Halg_2021}. Significantly higher forces can be generated by increasing the tip charge -- for the self-capacitance $C_\mathrm{self} = 4\pi\epsilon_0 R$ of a sphere with radius $R = \SI{50}{\nano\meter}$, a tip voltage $V_\mathrm{tip}=\SI{10}{\volt}$ induces $350$ elementary charges. We propose to implement the direct and parametric methods simultaneously to gain access to complementary information, i.e., the first and second derivatives of the tip-surface interaction energy, to improve the determination of the electrical fields.





To conclude, we have demonstrated strong (classical) parametric coupling between two membrane modes induced by the time-dependent voltage on a scanning tip. This achievement unlocks a large toolbox that vastly increases the range of possible experiments using such resonators. First, strong mode coupling was shown to be a useful way to avoid long transient dynamics and to obtain rapid resonator control~\cite{Okamoto_2014}. Second, our experiment serves an electrical test for parametric spin sensing experiments that will couple two resonator modes via coherently inverted nuclear spins~\cite{Dougherty_1996,Kosata_2020}. Third, our study motivates additional investigations of the membrane's electrical surface states, allowing deeper insights into the surface chemistry and the microscopic nature of non-contact friction~\cite{Halg_2021,heritier_2021}. Finally, with further improvements of the parametric coupling strength $g$, we could accomplish quantum-coherent state transfer between nondegenerate modes. The best dissipation-limited quantum coherence times for membrane resonators are on the order of milliseconds at \SI{4}{\kelvin}~\cite{Rossi_2018} and already exceed \SI{100}{\milli\second} at dilution refrigerator temperatures~\cite{Seis_2021}. This means that our experimental approach requires only a moderate improvement to overcome the decoherence stemming from dissipation, while decoherence caused by frequency noise is currently not well characterized in this system. Our scanning probe-based approach is, in addition, very flexible with regard to different modes of vibration, whose optimal coupling points (in real space) are located at different positions.



%
%


\begin{acknowledgments}
We gratefully acknowledge technical support by the mechanical workshop of D-PHYS at ETH Zurich. This work was supported by Swiss National Science Foundation (SNSF) through the National Center of Competence in Research in Quantum Science and Technology (NCCR QSIT), the Sinergia grant (CRSII5\_177198/1), and the project Grant 200020-178863, Danmarks Grundforskningsfond (DNRF) Center of Excellence Hy-Q, the European Research Council through the ERC Starting Grant Q-CEOM (grant agreement 638765), and the ERC Consolidator grant PHOQS (grant agreement 101002179), the DFG Heisenberg grant, as well as an ETH research grant (ETH-03 16-1).
\end{acknowledgments}

\providecommand{\noopsort}[1]{}\providecommand{\singleletter}[1]{#1}%


\begin{thebibliography}{36}%
	\makeatletter
	\providecommand \@ifxundefined [1]{%
		\@ifx{#1\undefined}
	}%
	\providecommand \@ifnum [1]{%
		\ifnum #1\expandafter \@firstoftwo
		\else \expandafter \@secondoftwo
		\fi
	}%
	\providecommand \@ifx [1]{%
		\ifx #1\expandafter \@firstoftwo
		\else \expandafter \@secondoftwo
		\fi
	}%
	\providecommand \natexlab [1]{#1}%
	\providecommand \enquote  [1]{``#1''}%
	\providecommand \bibnamefont  [1]{#1}%
	\providecommand \bibfnamefont [1]{#1}%
	\providecommand \citenamefont [1]{#1}%
	\providecommand \href@noop [0]{\@secondoftwo}%
	\providecommand \href [0]{\begingroup \@sanitize@url \@href}%
	\providecommand \@href[1]{\@@startlink{#1}\@@href}%
	\providecommand \@@href[1]{\endgroup#1\@@endlink}%
	\providecommand \@sanitize@url [0]{\catcode `\\12\catcode `\$12\catcode
		`\&12\catcode `\#12\catcode `\^12\catcode `\_12\catcode `\%12\relax}%
	\providecommand \@@startlink[1]{}%
	\providecommand \@@endlink[0]{}%
	\providecommand \url  [0]{\begingroup\@sanitize@url \@url }%
	\providecommand \@url [1]{\endgroup\@href {#1}{\urlprefix }}%
	\providecommand \urlprefix  [0]{URL }%
	\providecommand \Eprint [0]{\href }%
	\providecommand \doibase [0]{https://doi.org/}%
	\providecommand \selectlanguage [0]{\@gobble}%
	\providecommand \bibinfo  [0]{\@secondoftwo}%
	\providecommand \bibfield  [0]{\@secondoftwo}%
	\providecommand \translation [1]{[#1]}%
	\providecommand \BibitemOpen [0]{}%
	\providecommand \bibitemStop [0]{}%
	\providecommand \bibitemNoStop [0]{.\EOS\space}%
	\providecommand \EOS [0]{\spacefactor3000\relax}%
	\providecommand \BibitemShut  [1]{\csname bibitem#1\endcsname}%
	\let\auto@bib@innerbib\@empty
	\bibitem [{\citenamefont {Verbridge}\ \emph {et~al.}(2008)\citenamefont
		{Verbridge}, \citenamefont {Craighead},\ and\ \citenamefont
		{Parpia}}]{Verbridge_2008}%
	\BibitemOpen
	\bibfield  {author} {\bibinfo {author} {\bibfnamefont {S.~S.}\ \bibnamefont
			{Verbridge}}, \bibinfo {author} {\bibfnamefont {H.~G.}\ \bibnamefont
			{Craighead}},\ and\ \bibinfo {author} {\bibfnamefont {J.~M.}\ \bibnamefont
			{Parpia}},\ }\bibfield  {title} {\bibinfo {title} {A megahertz nanomechanical
			resonator with room temperature quality factor over a million},\ }\href
	{https://doi.org/10.1063/1.2822406} {\bibfield  {journal} {\bibinfo
			{journal} {Applied Physics Letters}\ }\textbf {\bibinfo {volume} {92}},\
		\bibinfo {pages} {013112} (\bibinfo {year} {2008})}\BibitemShut {NoStop}%
	\bibitem [{\citenamefont {Anetsberger}\ \emph {et~al.}(2010)\citenamefont
		{Anetsberger}, \citenamefont {Gavartin}, \citenamefont {Arcizet},
		\citenamefont {Unterreithmeier}, \citenamefont {Weig}, \citenamefont
		{Gorodetsky}, \citenamefont {Kotthaus},\ and\ \citenamefont
		{Kippenberg}}]{Anetsberger_2010}%
	\BibitemOpen
	\bibfield  {author} {\bibinfo {author} {\bibfnamefont {G.}~\bibnamefont
			{Anetsberger}}, \bibinfo {author} {\bibfnamefont {E.}~\bibnamefont
			{Gavartin}}, \bibinfo {author} {\bibfnamefont {O.}~\bibnamefont {Arcizet}},
		\bibinfo {author} {\bibfnamefont {Q.~P.}\ \bibnamefont {Unterreithmeier}},
		\bibinfo {author} {\bibfnamefont {E.~M.}\ \bibnamefont {Weig}}, \bibinfo
		{author} {\bibfnamefont {M.~L.}\ \bibnamefont {Gorodetsky}}, \bibinfo
		{author} {\bibfnamefont {J.~P.}\ \bibnamefont {Kotthaus}},\ and\ \bibinfo
		{author} {\bibfnamefont {T.~J.}\ \bibnamefont {Kippenberg}},\ }\bibfield
	{title} {\bibinfo {title} {Measuring nanomechanical motion with an
			imprecision below the standard quantum limit},\ }\href
	{https://doi.org/10.1103/PhysRevA.82.061804} {\bibfield  {journal} {\bibinfo
			{journal} {Phys. Rev. A}\ }\textbf {\bibinfo {volume} {82}},\ \bibinfo
		{pages} {061804(R)} (\bibinfo {year} {2010})}\BibitemShut {NoStop}%
	\bibitem [{\citenamefont {Reetz}\ \emph {et~al.}(2019)\citenamefont {Reetz},
		\citenamefont {Fischer}, \citenamefont {Assump{\c{c}}{\~{a}}o}, \citenamefont
		{McNally}, \citenamefont {Burns}, \citenamefont {Sankey},\ and\ \citenamefont
		{Regal}}]{Reetz_2019}%
	\BibitemOpen
	\bibfield  {author} {\bibinfo {author} {\bibfnamefont {C.}~\bibnamefont
			{Reetz}}, \bibinfo {author} {\bibfnamefont {R.}~\bibnamefont {Fischer}},
		\bibinfo {author} {\bibfnamefont {G.}~\bibnamefont {Assump{\c{c}}{\~{a}}o}},
		\bibinfo {author} {\bibfnamefont {D.}~\bibnamefont {McNally}}, \bibinfo
		{author} {\bibfnamefont {P.}~\bibnamefont {Burns}}, \bibinfo {author}
		{\bibfnamefont {J.}~\bibnamefont {Sankey}},\ and\ \bibinfo {author}
		{\bibfnamefont {C.}~\bibnamefont {Regal}},\ }\bibfield  {title} {\bibinfo
		{title} {Analysis of membrane phononic crystals with wide band gaps and
			low-mass defects},\ }\href {https://doi.org/10.1103/physrevapplied.12.044027}
	{\bibfield  {journal} {\bibinfo  {journal} {Physical Review Applied}\
		}\textbf {\bibinfo {volume} {12}},\ \bibinfo {pages} {044027} (\bibinfo
		{year} {2019})}\BibitemShut {NoStop}%
	\bibitem [{\citenamefont {Reinhardt}\ \emph {et~al.}(2016)\citenamefont
		{Reinhardt}, \citenamefont {M\"uller}, \citenamefont {Bourassa},\ and\
		\citenamefont {Sankey}}]{Reinhardt_2016}%
	\BibitemOpen
	\bibfield  {author} {\bibinfo {author} {\bibfnamefont {C.}~\bibnamefont
			{Reinhardt}}, \bibinfo {author} {\bibfnamefont {T.}~\bibnamefont {M\"uller}},
		\bibinfo {author} {\bibfnamefont {A.}~\bibnamefont {Bourassa}},\ and\
		\bibinfo {author} {\bibfnamefont {J.~C.}\ \bibnamefont {Sankey}},\ }\bibfield
	{title} {\bibinfo {title} {Ultralow-noise sin trampoline resonators for
			sensing and optomechanics},\ }\href
	{https://doi.org/10.1103/PhysRevX.6.021001} {\bibfield  {journal} {\bibinfo
			{journal} {Phys. Rev. X}\ }\textbf {\bibinfo {volume} {6}},\ \bibinfo {pages}
		{021001} (\bibinfo {year} {2016})}\BibitemShut {NoStop}%
	\bibitem [{\citenamefont {Tsaturyan}\ \emph {et~al.}(2017)\citenamefont
		{Tsaturyan}, \citenamefont {Barg}, \citenamefont {Polzik},\ and\
		\citenamefont {Schliesser}}]{Tsaturyan_2017}%
	\BibitemOpen
	\bibfield  {author} {\bibinfo {author} {\bibfnamefont {Y.}~\bibnamefont
			{Tsaturyan}}, \bibinfo {author} {\bibfnamefont {A.}~\bibnamefont {Barg}},
		\bibinfo {author} {\bibfnamefont {E.~S.}\ \bibnamefont {Polzik}},\ and\
		\bibinfo {author} {\bibfnamefont {A.}~\bibnamefont {Schliesser}},\ }\bibfield
	{title} {\bibinfo {title} {Ultracoherent nanomechanical resonators via soft
			clamping and dissipation dilution},\ }\href
	{https://doi.org/10.1038/nnano.2017.101} {\bibfield  {journal} {\bibinfo
			{journal} {Nature Nanotechnology}\ }\textbf {\bibinfo {volume} {12}},\
		\bibinfo {pages} {776} (\bibinfo {year} {2017})}\BibitemShut {NoStop}%
	\bibitem [{\citenamefont {Ghadimi}\ \emph {et~al.}(2018)\citenamefont
		{Ghadimi}, \citenamefont {Fedorov}, \citenamefont {Engelsen}, \citenamefont
		{Bereyhi}, \citenamefont {Schilling}, \citenamefont {Wilson},\ and\
		\citenamefont {Kippenberg}}]{Ghadimi_2018}%
	\BibitemOpen
	\bibfield  {author} {\bibinfo {author} {\bibfnamefont {A.~H.}\ \bibnamefont
			{Ghadimi}}, \bibinfo {author} {\bibfnamefont {S.~A.}\ \bibnamefont
			{Fedorov}}, \bibinfo {author} {\bibfnamefont {N.~J.}\ \bibnamefont
			{Engelsen}}, \bibinfo {author} {\bibfnamefont {M.~J.}\ \bibnamefont
			{Bereyhi}}, \bibinfo {author} {\bibfnamefont {R.}~\bibnamefont {Schilling}},
		\bibinfo {author} {\bibfnamefont {D.~J.}\ \bibnamefont {Wilson}},\ and\
		\bibinfo {author} {\bibfnamefont {T.~J.}\ \bibnamefont {Kippenberg}},\
	}\bibfield  {title} {\bibinfo {title} {Elastic strain engineering for
			ultralow mechanical dissipation},\ }\href
	{https://doi.org/10.1126/science.aar6939} {\bibfield  {journal} {\bibinfo
			{journal} {Science}\ }\textbf {\bibinfo {volume} {360}},\ \bibinfo {pages}
		{764} (\bibinfo {year} {2018})}\BibitemShut {NoStop}%
	\bibitem [{\citenamefont {Beccari}\ \emph
		{et~al.}(2021{\natexlab{a}})\citenamefont {Beccari}, \citenamefont {Bereyhi},
		\citenamefont {Groth}, \citenamefont {Fedorov}, \citenamefont {Arabmoheghi},
		\citenamefont {Engelsen},\ and\ \citenamefont
		{Kippenberg}}]{Beccari_2021_hierarchical}%
	\BibitemOpen
	\bibfield  {author} {\bibinfo {author} {\bibfnamefont {A.}~\bibnamefont
			{Beccari}}, \bibinfo {author} {\bibfnamefont {M.~J.}\ \bibnamefont
			{Bereyhi}}, \bibinfo {author} {\bibfnamefont {R.}~\bibnamefont {Groth}},
		\bibinfo {author} {\bibfnamefont {S.~A.}\ \bibnamefont {Fedorov}}, \bibinfo
		{author} {\bibfnamefont {A.}~\bibnamefont {Arabmoheghi}}, \bibinfo {author}
		{\bibfnamefont {N.~J.}\ \bibnamefont {Engelsen}},\ and\ \bibinfo {author}
		{\bibfnamefont {T.~J.}\ \bibnamefont {Kippenberg}},\ }\href@noop {} {\bibinfo
		{title} {Hierarchical tensile structures with ultralow mechanical
			dissipation}} (\bibinfo {year} {2021}{\natexlab{a}}),\ \Eprint
	{https://arxiv.org/abs/arXiv:2103.09785} {arXiv:2103.09785} \BibitemShut
	{NoStop}%
	\bibitem [{\citenamefont {Beccari}\ \emph
		{et~al.}(2021{\natexlab{b}})\citenamefont {Beccari}, \citenamefont {Visani},
		\citenamefont {Fedorov}, \citenamefont {Bereyhi}, \citenamefont {Boureau},
		\citenamefont {Engelsen},\ and\ \citenamefont {Kippenberg}}]{Beccari_2021}%
	\BibitemOpen
	\bibfield  {author} {\bibinfo {author} {\bibfnamefont {A.}~\bibnamefont
			{Beccari}}, \bibinfo {author} {\bibfnamefont {D.~A.}\ \bibnamefont {Visani}},
		\bibinfo {author} {\bibfnamefont {S.~A.}\ \bibnamefont {Fedorov}}, \bibinfo
		{author} {\bibfnamefont {M.~J.}\ \bibnamefont {Bereyhi}}, \bibinfo {author}
		{\bibfnamefont {V.}~\bibnamefont {Boureau}}, \bibinfo {author} {\bibfnamefont
			{N.~J.}\ \bibnamefont {Engelsen}},\ and\ \bibinfo {author} {\bibfnamefont
			{T.~J.}\ \bibnamefont {Kippenberg}},\ }\href@noop {} {\bibinfo {title}
		{Strained crystalline nanomechanical resonators with ultralow dissipation}}
	(\bibinfo {year} {2021}{\natexlab{b}}),\ \Eprint
	{https://arxiv.org/abs/arXiv:2107.02124} {arXiv:2107.02124} \BibitemShut
	{NoStop}%
	\bibitem [{\citenamefont {H\"alg}\ \emph {et~al.}(2021)\citenamefont {H\"alg},
		\citenamefont {Gisler}, \citenamefont {Tsaturyan}, \citenamefont {Catalini},
		\citenamefont {Grob}, \citenamefont {Krass}, \citenamefont {H\'eritier},
		\citenamefont {Mattiat}, \citenamefont {Thamm}, \citenamefont {Schirhagl},
		\citenamefont {Langman}, \citenamefont {Schliesser}, \citenamefont {Degen},\
		and\ \citenamefont {Eichler}}]{Halg_2021}%
	\BibitemOpen
	\bibfield  {author} {\bibinfo {author} {\bibfnamefont {D.}~\bibnamefont
			{H\"alg}}, \bibinfo {author} {\bibfnamefont {T.}~\bibnamefont {Gisler}},
		\bibinfo {author} {\bibfnamefont {Y.}~\bibnamefont {Tsaturyan}}, \bibinfo
		{author} {\bibfnamefont {L.}~\bibnamefont {Catalini}}, \bibinfo {author}
		{\bibfnamefont {U.}~\bibnamefont {Grob}}, \bibinfo {author} {\bibfnamefont
			{M.-D.}\ \bibnamefont {Krass}}, \bibinfo {author} {\bibfnamefont
			{M.}~\bibnamefont {H\'eritier}}, \bibinfo {author} {\bibfnamefont
			{H.}~\bibnamefont {Mattiat}}, \bibinfo {author} {\bibfnamefont {A.-K.}\
			\bibnamefont {Thamm}}, \bibinfo {author} {\bibfnamefont {R.}~\bibnamefont
			{Schirhagl}}, \bibinfo {author} {\bibfnamefont {E.~C.}\ \bibnamefont
			{Langman}}, \bibinfo {author} {\bibfnamefont {A.}~\bibnamefont {Schliesser}},
		\bibinfo {author} {\bibfnamefont {C.~L.}\ \bibnamefont {Degen}},\ and\
		\bibinfo {author} {\bibfnamefont {A.}~\bibnamefont {Eichler}},\ }\bibfield
	{title} {\bibinfo {title} {Membrane-based scanning force microscopy},\ }\href
	{https://doi.org/10.1103/PhysRevApplied.15.L021001} {\bibfield  {journal}
		{\bibinfo  {journal} {Phys. Rev. Applied}\ }\textbf {\bibinfo {volume}
			{15}},\ \bibinfo {pages} {L021001} (\bibinfo {year} {2021})}\BibitemShut
	{NoStop}%
	\bibitem [{\citenamefont {Peterson}\ \emph {et~al.}(2016)\citenamefont
		{Peterson}, \citenamefont {Purdy}, \citenamefont {Kampel}, \citenamefont
		{Andrews}, \citenamefont {Yu}, \citenamefont {Lehnert},\ and\ \citenamefont
		{Regal}}]{Peterson_2016}%
	\BibitemOpen
	\bibfield  {author} {\bibinfo {author} {\bibfnamefont {R.~W.}\ \bibnamefont
			{Peterson}}, \bibinfo {author} {\bibfnamefont {T.~P.}\ \bibnamefont {Purdy}},
		\bibinfo {author} {\bibfnamefont {N.~S.}\ \bibnamefont {Kampel}}, \bibinfo
		{author} {\bibfnamefont {R.~W.}\ \bibnamefont {Andrews}}, \bibinfo {author}
		{\bibfnamefont {P.-L.}\ \bibnamefont {Yu}}, \bibinfo {author} {\bibfnamefont
			{K.~W.}\ \bibnamefont {Lehnert}},\ and\ \bibinfo {author} {\bibfnamefont
			{C.~A.}\ \bibnamefont {Regal}},\ }\bibfield  {title} {\bibinfo {title} {Laser
			cooling of a micromechanical membrane to the quantum backaction limit},\
	}\href {https://doi.org/10.1103/PhysRevLett.116.063601} {\bibfield  {journal}
		{\bibinfo  {journal} {Phys. Rev. Lett.}\ }\textbf {\bibinfo {volume} {116}},\
		\bibinfo {pages} {063601} (\bibinfo {year} {2016})}\BibitemShut {NoStop}%
	\bibitem [{\citenamefont {Rossi}\ \emph {et~al.}(2018)\citenamefont {Rossi},
		\citenamefont {Mason}, \citenamefont {Chen}, \citenamefont {Tsaturyan},\ and\
		\citenamefont {Schliesser}}]{Rossi_2018}%
	\BibitemOpen
	\bibfield  {author} {\bibinfo {author} {\bibfnamefont {M.}~\bibnamefont
			{Rossi}}, \bibinfo {author} {\bibfnamefont {D.}~\bibnamefont {Mason}},
		\bibinfo {author} {\bibfnamefont {J.}~\bibnamefont {Chen}}, \bibinfo {author}
		{\bibfnamefont {Y.}~\bibnamefont {Tsaturyan}},\ and\ \bibinfo {author}
		{\bibfnamefont {A.}~\bibnamefont {Schliesser}},\ }\bibfield  {title}
	{\bibinfo {title} {Measurement-based quantum control of mechanical motion},\
	}\href {https://doi.org/10.1038/s41586-018-0643-8} {\bibfield  {journal}
		{\bibinfo  {journal} {Nature}\ }\textbf {\bibinfo {volume} {563}},\ \bibinfo
		{pages} {53} (\bibinfo {year} {2018})}\BibitemShut {NoStop}%
	\bibitem [{\citenamefont {Mason}\ \emph {et~al.}(2019)\citenamefont {Mason},
		\citenamefont {Chen}, \citenamefont {Rossi}, \citenamefont {Tsaturyan},\ and\
		\citenamefont {Schliesser}}]{mason_continuous_2019}%
	\BibitemOpen
	\bibfield  {author} {\bibinfo {author} {\bibfnamefont {D.}~\bibnamefont
			{Mason}}, \bibinfo {author} {\bibfnamefont {J.}~\bibnamefont {Chen}},
		\bibinfo {author} {\bibfnamefont {M.}~\bibnamefont {Rossi}}, \bibinfo
		{author} {\bibfnamefont {Y.}~\bibnamefont {Tsaturyan}},\ and\ \bibinfo
		{author} {\bibfnamefont {A.}~\bibnamefont {Schliesser}},\ }\bibfield  {title}
	{\bibinfo {title} {Continuous force and displacement measurement below the
			standard quantum limit},\ }\href {https://doi.org/10.1038/s41567-019-0533-5}
	{\bibfield  {journal} {\bibinfo  {journal} {Nature Physics}\ }\textbf
		{\bibinfo {volume} {15}},\ \bibinfo {pages} {745} (\bibinfo {year}
		{2019})}\BibitemShut {NoStop}%
	\bibitem [{\citenamefont {Page}\ \emph {et~al.}(2021)\citenamefont {Page},
		\citenamefont {Goryachev}, \citenamefont {Miao}, \citenamefont {Chen},
		\citenamefont {Ma}, \citenamefont {Mason}, \citenamefont {Rossi},
		\citenamefont {Blair}, \citenamefont {Ju}, \citenamefont {Blair},
		\citenamefont {Schliesser}, \citenamefont {Tobar},\ and\ \citenamefont
		{Zhao}}]{Page_2021_gravitational}%
	\BibitemOpen
	\bibfield  {author} {\bibinfo {author} {\bibfnamefont {M.~A.}\ \bibnamefont
			{Page}}, \bibinfo {author} {\bibfnamefont {M.}~\bibnamefont {Goryachev}},
		\bibinfo {author} {\bibfnamefont {H.}~\bibnamefont {Miao}}, \bibinfo {author}
		{\bibfnamefont {Y.}~\bibnamefont {Chen}}, \bibinfo {author} {\bibfnamefont
			{Y.}~\bibnamefont {Ma}}, \bibinfo {author} {\bibfnamefont {D.}~\bibnamefont
			{Mason}}, \bibinfo {author} {\bibfnamefont {M.}~\bibnamefont {Rossi}},
		\bibinfo {author} {\bibfnamefont {C.~D.}\ \bibnamefont {Blair}}, \bibinfo
		{author} {\bibfnamefont {L.}~\bibnamefont {Ju}}, \bibinfo {author}
		{\bibfnamefont {D.~G.}\ \bibnamefont {Blair}}, \bibinfo {author}
		{\bibfnamefont {A.}~\bibnamefont {Schliesser}}, \bibinfo {author}
		{\bibfnamefont {M.~E.}\ \bibnamefont {Tobar}},\ and\ \bibinfo {author}
		{\bibfnamefont {C.}~\bibnamefont {Zhao}},\ }\bibfield  {title} {\bibinfo
		{title} {Gravitational wave detectors with broadband high frequency
			sensitivity},\ }\href {https://doi.org/10.1038/s42005-021-00526-2} {\bibfield
		{journal} {\bibinfo  {journal} {Communications Physics}\ }\textbf {\bibinfo
			{volume} {4}} (\bibinfo {year} {2021})}\BibitemShut {NoStop}%
	\bibitem [{\citenamefont {Bagci}\ \emph {et~al.}(2014)\citenamefont {Bagci},
		\citenamefont {Simonsen}, \citenamefont {Schmid}, \citenamefont {Villanueva},
		\citenamefont {Zeuthen}, \citenamefont {Appel}, \citenamefont {Taylor},
		\citenamefont {S{\o}rensen}, \citenamefont {Usami}, \citenamefont
		{Schliesser},\ and\ \citenamefont {Polzik}}]{Bagci_2014}%
	\BibitemOpen
	\bibfield  {author} {\bibinfo {author} {\bibfnamefont {T.}~\bibnamefont
			{Bagci}}, \bibinfo {author} {\bibfnamefont {A.}~\bibnamefont {Simonsen}},
		\bibinfo {author} {\bibfnamefont {S.}~\bibnamefont {Schmid}}, \bibinfo
		{author} {\bibfnamefont {L.~G.}\ \bibnamefont {Villanueva}}, \bibinfo
		{author} {\bibfnamefont {E.}~\bibnamefont {Zeuthen}}, \bibinfo {author}
		{\bibfnamefont {J.}~\bibnamefont {Appel}}, \bibinfo {author} {\bibfnamefont
			{J.~M.}\ \bibnamefont {Taylor}}, \bibinfo {author} {\bibfnamefont
			{A.}~\bibnamefont {S{\o}rensen}}, \bibinfo {author} {\bibfnamefont
			{K.}~\bibnamefont {Usami}}, \bibinfo {author} {\bibfnamefont
			{A.}~\bibnamefont {Schliesser}},\ and\ \bibinfo {author} {\bibfnamefont
			{E.~S.}\ \bibnamefont {Polzik}},\ }\bibfield  {title} {\bibinfo {title}
		{Optical detection of radio waves through a nanomechanical transducer},\
	}\href {https://doi.org/10.1038/nature13029} {\bibfield  {journal} {\bibinfo
			{journal} {Nature}\ }\textbf {\bibinfo {volume} {507}},\ \bibinfo {pages}
		{81} (\bibinfo {year} {2014})}\BibitemShut {NoStop}%
	\bibitem [{\citenamefont {Andrews}\ \emph {et~al.}(2014)\citenamefont
		{Andrews}, \citenamefont {Peterson}, \citenamefont {Purdy}, \citenamefont
		{Cicak}, \citenamefont {Simmonds}, \citenamefont {Regal},\ and\ \citenamefont
		{Lehnert}}]{Andrews_2014}%
	\BibitemOpen
	\bibfield  {author} {\bibinfo {author} {\bibfnamefont {R.~W.}\ \bibnamefont
			{Andrews}}, \bibinfo {author} {\bibfnamefont {R.~W.}\ \bibnamefont
			{Peterson}}, \bibinfo {author} {\bibfnamefont {T.~P.}\ \bibnamefont {Purdy}},
		\bibinfo {author} {\bibfnamefont {K.}~\bibnamefont {Cicak}}, \bibinfo
		{author} {\bibfnamefont {R.~W.}\ \bibnamefont {Simmonds}}, \bibinfo {author}
		{\bibfnamefont {C.~A.}\ \bibnamefont {Regal}},\ and\ \bibinfo {author}
		{\bibfnamefont {K.~W.}\ \bibnamefont {Lehnert}},\ }\bibfield  {title}
	{\bibinfo {title} {Bidirectional and efficient conversion between microwave
			and optical light},\ }\href {https://doi.org/10.1038/nphys2911} {\bibfield
		{journal} {\bibinfo  {journal} {Nature Physics}\ }\textbf {\bibinfo {volume}
			{10}},\ \bibinfo {pages} {321} (\bibinfo {year} {2014})}\BibitemShut
	{NoStop}%
	\bibitem [{\citenamefont {Fischer}\ \emph {et~al.}(2019)\citenamefont
		{Fischer}, \citenamefont {McNally}, \citenamefont {Reetz}, \citenamefont
		{Assump{\c{c}}{\~{a}}o}, \citenamefont {Knief}, \citenamefont {Lin},\ and\
		\citenamefont {Regal}}]{Fischer_2019}%
	\BibitemOpen
	\bibfield  {author} {\bibinfo {author} {\bibfnamefont {R.}~\bibnamefont
			{Fischer}}, \bibinfo {author} {\bibfnamefont {D.~P.}\ \bibnamefont
			{McNally}}, \bibinfo {author} {\bibfnamefont {C.}~\bibnamefont {Reetz}},
		\bibinfo {author} {\bibfnamefont {G.~G.~T.}\ \bibnamefont
			{Assump{\c{c}}{\~{a}}o}}, \bibinfo {author} {\bibfnamefont {T.}~\bibnamefont
			{Knief}}, \bibinfo {author} {\bibfnamefont {Y.}~\bibnamefont {Lin}},\ and\
		\bibinfo {author} {\bibfnamefont {C.~A.}\ \bibnamefont {Regal}},\ }\bibfield
	{title} {\bibinfo {title} {Spin detection with a micromechanical trampoline:
			towards magnetic resonance microscopy harnessing cavity optomechanics},\
	}\href {https://doi.org/10.1088/1367-2630/ab117a} {\bibfield  {journal}
		{\bibinfo  {journal} {New Journal of Physics}\ }\textbf {\bibinfo {volume}
			{21}},\ \bibinfo {pages} {043049} (\bibinfo {year} {2019})}\BibitemShut
	{NoStop}%
	\bibitem [{\citenamefont {Ko\ifmmode~\check{s}\else \v{s}\fi{}ata}\ \emph
		{et~al.}(2020)\citenamefont {Ko\ifmmode~\check{s}\else \v{s}\fi{}ata},
		\citenamefont {Zilberberg}, \citenamefont {Degen}, \citenamefont {Chitra},\
		and\ \citenamefont {Eichler}}]{Kosata_2020}%
	\BibitemOpen
	\bibfield  {author} {\bibinfo {author} {\bibfnamefont {J.}~\bibnamefont
			{Ko\ifmmode~\check{s}\else \v{s}\fi{}ata}}, \bibinfo {author} {\bibfnamefont
			{O.}~\bibnamefont {Zilberberg}}, \bibinfo {author} {\bibfnamefont {C.~L.}\
			\bibnamefont {Degen}}, \bibinfo {author} {\bibfnamefont {R.}~\bibnamefont
			{Chitra}},\ and\ \bibinfo {author} {\bibfnamefont {A.}~\bibnamefont
			{Eichler}},\ }\bibfield  {title} {\bibinfo {title} {Spin detection via
			parametric frequency conversion in a membrane resonator},\ }\href
	{https://doi.org/10.1103/PhysRevApplied.14.014042} {\bibfield  {journal}
		{\bibinfo  {journal} {Phys. Rev. Applied}\ }\textbf {\bibinfo {volume}
			{14}},\ \bibinfo {pages} {014042} (\bibinfo {year} {2020})}\BibitemShut
	{NoStop}%
	\bibitem [{\citenamefont {Boyd}(2020)}]{Boyd_2020}%
	\BibitemOpen
	\bibfield  {author} {\bibinfo {author} {\bibfnamefont {R.~W.}\ \bibnamefont
			{Boyd}},\ }\href@noop {} {\emph {\bibinfo {title} {Nonlinear optics}}}\
	(\bibinfo  {publisher} {Academic press},\ \bibinfo {year} {2020})\BibitemShut
	{NoStop}%
	\bibitem [{\citenamefont {Dougherty}\ \emph {et~al.}(1996)\citenamefont
		{Dougherty}, \citenamefont {Bruland}, \citenamefont {Garbini},\ and\
		\citenamefont {Sidles}}]{Dougherty_1996}%
	\BibitemOpen
	\bibfield  {author} {\bibinfo {author} {\bibfnamefont {W.~M.}\ \bibnamefont
			{Dougherty}}, \bibinfo {author} {\bibfnamefont {K.~J.}\ \bibnamefont
			{Bruland}}, \bibinfo {author} {\bibfnamefont {J.~L.}\ \bibnamefont
			{Garbini}},\ and\ \bibinfo {author} {\bibfnamefont {J.~A.}\ \bibnamefont
			{Sidles}},\ }\bibfield  {title} {\bibinfo {title} {Detection of ac magnetic
			signals by parametric mode coupling in a mechanical oscillator},\ }\href
	{https://doi.org/10.1088/0957-0233/7/12/007} {\bibfield  {journal} {\bibinfo
			{journal} {Meas. Sci. Technol.}\ }\textbf {\bibinfo {volume} {7}},\ \bibinfo
		{pages} {1733–1739} (\bibinfo {year} {1996})}\BibitemShut {NoStop}%
	\bibitem [{\citenamefont {Faust}\ \emph {et~al.}(2013)\citenamefont {Faust},
		\citenamefont {Rieger}, \citenamefont {Seitner}, \citenamefont {Kotthaus},\
		and\ \citenamefont {Weig}}]{Faust_2013}%
	\BibitemOpen
	\bibfield  {author} {\bibinfo {author} {\bibfnamefont {T.}~\bibnamefont
			{Faust}}, \bibinfo {author} {\bibfnamefont {J.}~\bibnamefont {Rieger}},
		\bibinfo {author} {\bibfnamefont {M.~J.}\ \bibnamefont {Seitner}}, \bibinfo
		{author} {\bibfnamefont {J.~P.}\ \bibnamefont {Kotthaus}},\ and\ \bibinfo
		{author} {\bibfnamefont {E.~M.}\ \bibnamefont {Weig}},\ }\bibfield  {title}
	{\bibinfo {title} {Coherent control of a classical nanomechanical two-level
			system},\ }\href {https://doi.org/10.1038/nphys2666} {\bibfield  {journal}
		{\bibinfo  {journal} {Nature Physics}\ }\textbf {\bibinfo {volume} {9}},\
		\bibinfo {pages} {485} (\bibinfo {year} {2013})}\BibitemShut {NoStop}%
	\bibitem [{\citenamefont {Okamoto}\ \emph
		{et~al.}(2013{\natexlab{a}})\citenamefont {Okamoto}, \citenamefont
		{Gourgout}, \citenamefont {Chang}, \citenamefont {Onomitsu}, \citenamefont
		{Mahboob}, \citenamefont {Chang},\ and\ \citenamefont
		{Yamaguchi}}]{Okamoto_2013}%
	\BibitemOpen
	\bibfield  {author} {\bibinfo {author} {\bibfnamefont {H.}~\bibnamefont
			{Okamoto}}, \bibinfo {author} {\bibfnamefont {A.}~\bibnamefont {Gourgout}},
		\bibinfo {author} {\bibfnamefont {C.-Y.}\ \bibnamefont {Chang}}, \bibinfo
		{author} {\bibfnamefont {K.}~\bibnamefont {Onomitsu}}, \bibinfo {author}
		{\bibfnamefont {I.}~\bibnamefont {Mahboob}}, \bibinfo {author} {\bibfnamefont
			{E.~Y.}\ \bibnamefont {Chang}},\ and\ \bibinfo {author} {\bibfnamefont
			{H.}~\bibnamefont {Yamaguchi}},\ }\bibfield  {title} {\bibinfo {title}
		{Coherent phonon manipulation in coupled mechanical resonators},\ }\href
	{https://doi.org/10.1038/nphys2665} {\bibfield  {journal} {\bibinfo
			{journal} {Nat. Phys.}\ }\textbf {\bibinfo {volume} {9}},\ \bibinfo {pages}
		{480} (\bibinfo {year} {2013}{\natexlab{a}})}\BibitemShut {NoStop}%
	\bibitem [{\citenamefont {Seis}\ \emph {et~al.}(2021)\citenamefont {Seis},
		\citenamefont {Capelle}, \citenamefont {Langman}, \citenamefont {Saarinen},
		\citenamefont {Planz},\ and\ \citenamefont {Schliesser}}]{Seis_2021}%
	\BibitemOpen
	\bibfield  {author} {\bibinfo {author} {\bibfnamefont {Y.}~\bibnamefont
			{Seis}}, \bibinfo {author} {\bibfnamefont {T.}~\bibnamefont {Capelle}},
		\bibinfo {author} {\bibfnamefont {E.}~\bibnamefont {Langman}}, \bibinfo
		{author} {\bibfnamefont {S.}~\bibnamefont {Saarinen}}, \bibinfo {author}
		{\bibfnamefont {E.}~\bibnamefont {Planz}},\ and\ \bibinfo {author}
		{\bibfnamefont {A.}~\bibnamefont {Schliesser}},\ }\href@noop {} {\bibinfo
		{title} {Ground state cooling of an ultracoherent electromechanical system}}
	(\bibinfo {year} {2021}),\ \Eprint {https://arxiv.org/abs/arXiv:2107.05552}
	{arXiv:2107.05552} \BibitemShut {NoStop}%
	\bibitem [{\citenamefont {Catalini}\ \emph {et~al.}(2020)\citenamefont
		{Catalini}, \citenamefont {Tsaturyan},\ and\ \citenamefont
		{Schliesser}}]{Catalini_2020}%
	\BibitemOpen
	\bibfield  {author} {\bibinfo {author} {\bibfnamefont {L.}~\bibnamefont
			{Catalini}}, \bibinfo {author} {\bibfnamefont {Y.}~\bibnamefont
			{Tsaturyan}},\ and\ \bibinfo {author} {\bibfnamefont {A.}~\bibnamefont
			{Schliesser}},\ }\bibfield  {title} {\bibinfo {title} {Soft-clamped phononic
			dimers for mechanical sensing and transduction},\ }\href
	{https://doi.org/10.1103/PhysRevApplied.14.014041} {\bibfield  {journal}
		{\bibinfo  {journal} {Phys. Rev. Applied}\ }\textbf {\bibinfo {volume}
			{14}},\ \bibinfo {pages} {014041} (\bibinfo {year} {2020})}\BibitemShut
	{NoStop}%
	\bibitem [{\citenamefont {Unterreithmeier}\ \emph {et~al.}(2009)\citenamefont
		{Unterreithmeier}, \citenamefont {Weig},\ and\ \citenamefont
		{Kotthaus}}]{Unterreithmeier2009}%
	\BibitemOpen
	\bibfield  {author} {\bibinfo {author} {\bibfnamefont {Q.~P.}\ \bibnamefont
			{Unterreithmeier}}, \bibinfo {author} {\bibfnamefont {E.~M.}\ \bibnamefont
			{Weig}},\ and\ \bibinfo {author} {\bibfnamefont {J.~P.}\ \bibnamefont
			{Kotthaus}},\ }\bibfield  {title} {\bibinfo {title} {Universal transduction
			scheme for nanomechanical systems based on dielectric forces},\ }\href
	{https://doi.org/10.1038/nature07932} {\bibfield  {journal} {\bibinfo
			{journal} {Nature}\ }\textbf {\bibinfo {volume} {458}},\ \bibinfo {pages}
		{1001} (\bibinfo {year} {2009})}\BibitemShut {NoStop}%
	\bibitem [{\citenamefont {Camp}\ \emph {et~al.}(1991)\citenamefont {Camp},
		\citenamefont {Darling},\ and\ \citenamefont
		{Brown}}]{camp_macroscopic_1991}%
	\BibitemOpen
	\bibfield  {author} {\bibinfo {author} {\bibfnamefont {J.~B.}\ \bibnamefont
			{Camp}}, \bibinfo {author} {\bibfnamefont {T.~W.}\ \bibnamefont {Darling}},\
		and\ \bibinfo {author} {\bibfnamefont {R.~E.}\ \bibnamefont {Brown}},\
	}\bibfield  {title} {\bibinfo {title} {Macroscopic variations of surface
			potentials of conductors},\ }\href {https://doi.org/10.1063/1.347601}
	{\bibfield  {journal} {\bibinfo  {journal} {Journal of Applied Physics}\
		}\textbf {\bibinfo {volume} {69}},\ \bibinfo {pages} {7126} (\bibinfo {year}
		{1991})}\BibitemShut {NoStop}%
	\bibitem [{\citenamefont {Burnham}\ \emph {et~al.}(1992)\citenamefont
		{Burnham}, \citenamefont {Colton},\ and\ \citenamefont
		{Pollock}}]{burnham_work-function_1992}%
	\BibitemOpen
	\bibfield  {author} {\bibinfo {author} {\bibfnamefont {N.~A.}\ \bibnamefont
			{Burnham}}, \bibinfo {author} {\bibfnamefont {R.~J.}\ \bibnamefont
			{Colton}},\ and\ \bibinfo {author} {\bibfnamefont {H.~M.}\ \bibnamefont
			{Pollock}},\ }\bibfield  {title} {\bibinfo {title} {Work-function
			anisotropies as an origin of long-range surface forces},\ }\href
	{https://doi.org/10.1103/physrevlett.69.144} {\bibfield  {journal} {\bibinfo
			{journal} {Physical Review Letters}\ }\textbf {\bibinfo {volume} {69}},\
		\bibinfo {pages} {144} (\bibinfo {year} {1992})}\BibitemShut {NoStop}%
	\bibitem [{\citenamefont {Rossi}\ and\ \citenamefont
		{Opat}(1992)}]{rossi_observations_1992}%
	\BibitemOpen
	\bibfield  {author} {\bibinfo {author} {\bibfnamefont {F.}~\bibnamefont
			{Rossi}}\ and\ \bibinfo {author} {\bibfnamefont {G.~I.}\ \bibnamefont
			{Opat}},\ }\bibfield  {title} {\bibinfo {title} {Observations of the effects
			of adsorbates on patch potentials},\ }\href
	{https://doi.org/10.1088/0022-3727/25/9/012} {\bibfield  {journal} {\bibinfo
			{journal} {Journal of Physics D: Applied Physics}\ }\textbf {\bibinfo
			{volume} {25}},\ \bibinfo {pages} {1349} (\bibinfo {year}
		{1992})}\BibitemShut {NoStop}%
	\bibitem [{\citenamefont {Speake}\ and\ \citenamefont
		{Trenkel}(2003)}]{speake_forces_2003}%
	\BibitemOpen
	\bibfield  {author} {\bibinfo {author} {\bibfnamefont {C.~C.}\ \bibnamefont
			{Speake}}\ and\ \bibinfo {author} {\bibfnamefont {C.}~\bibnamefont
			{Trenkel}},\ }\bibfield  {title} {\bibinfo {title} {Forces between conducting
			surfaces due to spatial variations of surface potential},\ }\href
	{https://doi.org/10.1103/physrevlett.90.160403} {\bibfield  {journal}
		{\bibinfo  {journal} {Physical Review Letters}\ }\textbf {\bibinfo {volume}
			{90}},\ \bibinfo {pages} {160403} (\bibinfo {year} {2003})}\BibitemShut
	{NoStop}%
	\bibitem [{\citenamefont {Gaillard}\ \emph {et~al.}(2006)\citenamefont
		{Gaillard}, \citenamefont {Gros-Jean}, \citenamefont {Mariolle},
		\citenamefont {Bertin},\ and\ \citenamefont {Bsiesy}}]{gaillard_method_2006}%
	\BibitemOpen
	\bibfield  {author} {\bibinfo {author} {\bibfnamefont {N.}~\bibnamefont
			{Gaillard}}, \bibinfo {author} {\bibfnamefont {M.}~\bibnamefont {Gros-Jean}},
		\bibinfo {author} {\bibfnamefont {D.}~\bibnamefont {Mariolle}}, \bibinfo
		{author} {\bibfnamefont {F.}~\bibnamefont {Bertin}},\ and\ \bibinfo {author}
		{\bibfnamefont {A.}~\bibnamefont {Bsiesy}},\ }\bibfield  {title} {\bibinfo
		{title} {Method to assess the grain crystallographic orientation with a
			submicronic spatial resolution using kelvin probe force microscope},\ }\href
	{https://doi.org/10.1063/1.2359297} {\bibfield  {journal} {\bibinfo
			{journal} {Applied Physics Letters}\ }\textbf {\bibinfo {volume} {89}},\
		\bibinfo {pages} {154101} (\bibinfo {year} {2006})}\BibitemShut {NoStop}%
	\bibitem [{\citenamefont {Robertson}\ \emph {et~al.}(2006)\citenamefont
		{Robertson}, \citenamefont {Blackwood}, \citenamefont {Buchman},
		\citenamefont {Byer}, \citenamefont {Camp}, \citenamefont {Gill},
		\citenamefont {Hanson}, \citenamefont {Williams},\ and\ \citenamefont
		{Zhou}}]{robertson_kelvin_2006}%
	\BibitemOpen
	\bibfield  {author} {\bibinfo {author} {\bibfnamefont {N.~A.}\ \bibnamefont
			{Robertson}}, \bibinfo {author} {\bibfnamefont {J.~R.}\ \bibnamefont
			{Blackwood}}, \bibinfo {author} {\bibfnamefont {S.}~\bibnamefont {Buchman}},
		\bibinfo {author} {\bibfnamefont {R.~L.}\ \bibnamefont {Byer}}, \bibinfo
		{author} {\bibfnamefont {J.}~\bibnamefont {Camp}}, \bibinfo {author}
		{\bibfnamefont {D.}~\bibnamefont {Gill}}, \bibinfo {author} {\bibfnamefont
			{J.}~\bibnamefont {Hanson}}, \bibinfo {author} {\bibfnamefont
			{S.}~\bibnamefont {Williams}},\ and\ \bibinfo {author} {\bibfnamefont
			{P.}~\bibnamefont {Zhou}},\ }\bibfield  {title} {\bibinfo {title} {Kelvin
			probe measurements: investigations of the patch effect with applications to
			{ST}-7 and {LISA}},\ }\href {https://doi.org/10.1088/0264-9381/23/7/026}
	{\bibfield  {journal} {\bibinfo  {journal} {Classical and Quantum Gravity}\
		}\textbf {\bibinfo {volume} {23}},\ \bibinfo {pages} {2665} (\bibinfo {year}
		{2006})}\BibitemShut {NoStop}%
	\bibitem [{\citenamefont {Z{\'{u}}{\~{n}}iga-P{\'{e}}rez}\ \emph
		{et~al.}(2005)\citenamefont {Z{\'{u}}{\~{n}}iga-P{\'{e}}rez}, \citenamefont
		{Mu{\~{n}}oz-Sanjos{\'{e}}}, \citenamefont {Palacios-Lid{\'{o}}n},\ and\
		\citenamefont {Colchero}}]{zuniga_polarity_2005}%
	\BibitemOpen
	\bibfield  {author} {\bibinfo {author} {\bibfnamefont {J.}~\bibnamefont
			{Z{\'{u}}{\~{n}}iga-P{\'{e}}rez}}, \bibinfo {author} {\bibfnamefont
			{V.}~\bibnamefont {Mu{\~{n}}oz-Sanjos{\'{e}}}}, \bibinfo {author}
		{\bibfnamefont {E.}~\bibnamefont {Palacios-Lid{\'{o}}n}},\ and\ \bibinfo
		{author} {\bibfnamefont {J.}~\bibnamefont {Colchero}},\ }\bibfield  {title}
	{\bibinfo {title} {Polarity effects on zno films grown along the nonpolar
			$[11\overline{2}0]$ direction},\ }\href
	{https://doi.org/10.1103/physrevlett.95.226105} {\bibfield  {journal}
		{\bibinfo  {journal} {Physical Review Letters}\ }\textbf {\bibinfo {volume}
			{95}},\ \bibinfo {pages} {226105} (\bibinfo {year} {2005})}\BibitemShut
	{NoStop}%
	\bibitem [{\citenamefont {Héritier}\ \emph {et~al.}(2021)\citenamefont
		{Héritier}, \citenamefont {Pachlatko}, \citenamefont {Tao}, \citenamefont
		{Abendroth}, \citenamefont {Degen},\ and\ \citenamefont
		{Eichler}}]{heritier_2021}%
	\BibitemOpen
	\bibfield  {author} {\bibinfo {author} {\bibfnamefont {M.}~\bibnamefont
			{Héritier}}, \bibinfo {author} {\bibfnamefont {R.}~\bibnamefont
			{Pachlatko}}, \bibinfo {author} {\bibfnamefont {Y.}~\bibnamefont {Tao}},
		\bibinfo {author} {\bibfnamefont {J.~M.}\ \bibnamefont {Abendroth}}, \bibinfo
		{author} {\bibfnamefont {C.~L.}\ \bibnamefont {Degen}},\ and\ \bibinfo
		{author} {\bibfnamefont {A.}~\bibnamefont {Eichler}},\ }\href@noop {}
	{\bibinfo {title} {Spatially resolved surface dissipation over metal and
			dielectric substrates}} (\bibinfo {year} {2021}),\ \Eprint
	{https://arxiv.org/abs/arXiv:2104.02437} {arXiv:2104.02437} \BibitemShut
	{NoStop}%
	\bibitem [{Sup()}]{Supplement}%
	\BibitemOpen
	\href@noop {} {\bibinfo {title} {See supplemental material at [url will be
			inserted by publisher] for additional discussions on the surface charge, data
			of a second device, the derivation of the coupling process and driving at the
			frequency sum.}}\BibitemShut {Stop}%
	\bibitem [{\citenamefont {Frimmer}\ and\ \citenamefont
		{Novotny}(2014)}]{Frimmer_2014}%
	\BibitemOpen
	\bibfield  {author} {\bibinfo {author} {\bibfnamefont {M.}~\bibnamefont
			{Frimmer}}\ and\ \bibinfo {author} {\bibfnamefont {L.}~\bibnamefont
			{Novotny}},\ }\bibfield  {title} {\bibinfo {title} {The classical bloch
			equations},\ }\href {https://doi.org/10.1119/1.4878621} {\bibfield  {journal}
		{\bibinfo  {journal} {Am. J. Phys}\ }\textbf {\bibinfo {volume} {82}},\
		\bibinfo {pages} {947–954} (\bibinfo {year} {2014})}\BibitemShut {NoStop}%
	\bibitem [{\citenamefont {Okamoto}\ \emph
		{et~al.}(2013{\natexlab{b}})\citenamefont {Okamoto}, \citenamefont
		{Gourgout}, \citenamefont {Chang}, \citenamefont {Onomitsu}, \citenamefont
		{Mahboob}, \citenamefont {Chang},\ and\ \citenamefont
		{Yamaguchi}}]{Okamoto_2013_NP}%
	\BibitemOpen
	\bibfield  {author} {\bibinfo {author} {\bibfnamefont {H.}~\bibnamefont
			{Okamoto}}, \bibinfo {author} {\bibfnamefont {A.}~\bibnamefont {Gourgout}},
		\bibinfo {author} {\bibfnamefont {C.-Y.}\ \bibnamefont {Chang}}, \bibinfo
		{author} {\bibfnamefont {K.}~\bibnamefont {Onomitsu}}, \bibinfo {author}
		{\bibfnamefont {I.}~\bibnamefont {Mahboob}}, \bibinfo {author} {\bibfnamefont
			{E.~Y.}\ \bibnamefont {Chang}},\ and\ \bibinfo {author} {\bibfnamefont
			{H.}~\bibnamefont {Yamaguchi}},\ }\bibfield  {title} {\bibinfo {title}
		{Coherent phonon manipulation in coupled mechanical resonators},\ }\href
	{https://doi.org/10.1038/nphys2665} {\bibfield  {journal} {\bibinfo
			{journal} {Nature Physics}\ }\textbf {\bibinfo {volume} {9}},\ \bibinfo
		{pages} {480} (\bibinfo {year} {2013}{\natexlab{b}})}\BibitemShut {NoStop}%
	\bibitem [{\citenamefont {Okamoto}\ \emph {et~al.}(2014)\citenamefont
		{Okamoto}, \citenamefont {Mahboob}, \citenamefont {Onomitsu},\ and\
		\citenamefont {Yamaguchi}}]{Okamoto_2014}%
	\BibitemOpen
	\bibfield  {author} {\bibinfo {author} {\bibfnamefont {H.}~\bibnamefont
			{Okamoto}}, \bibinfo {author} {\bibfnamefont {I.}~\bibnamefont {Mahboob}},
		\bibinfo {author} {\bibfnamefont {K.}~\bibnamefont {Onomitsu}},\ and\
		\bibinfo {author} {\bibfnamefont {H.}~\bibnamefont {Yamaguchi}},\ }\bibfield
	{title} {\bibinfo {title} {Rapid switching in high-q mechanical resonators},\
	}\href {https://doi.org/10.1063/1.4894417} {\bibfield  {journal} {\bibinfo
			{journal} {Applied Physics Letters}\ }\textbf {\bibinfo {volume} {105}},\
		\bibinfo {pages} {083114} (\bibinfo {year} {2014})}\BibitemShut {NoStop}%
\end{thebibliography}
\end{document}


\title{\textbf{Supplemental Material for:}\\ Strong parametric coupling between two ultra-coherent membrane modes}
\author{David H\"alg}
\affiliation{\affilETH}
\author{Thomas Gisler}
\affiliation{\affilETH}
\author{Eric C. Langman}
\affiliation{\affilNBI}
\affiliation{\affilNBIHYQ}
\author{Shobhna Misra}
\affiliation{\affilETH}
\author{Oded Zilberberg}
\affiliation{\ITP}
\affiliation{\affilKonstanz}
\author{Albert Schliesser}
\affiliation{\affilNBI}
\affiliation{\affilNBIHYQ}
\author{Christian L. Degen}
\affiliation{\affilETH}
\affiliation{\affilETHQuantum}
\author{Alexander Eichler}
\email[Corresponding author: ]{eichlera@ethz.ch}
\affiliation{\affilETH}

\maketitle

\spacing{1.5}

\section{Charge investigations}

    \begin{figure*}
    \includegraphics[width=\textwidth]{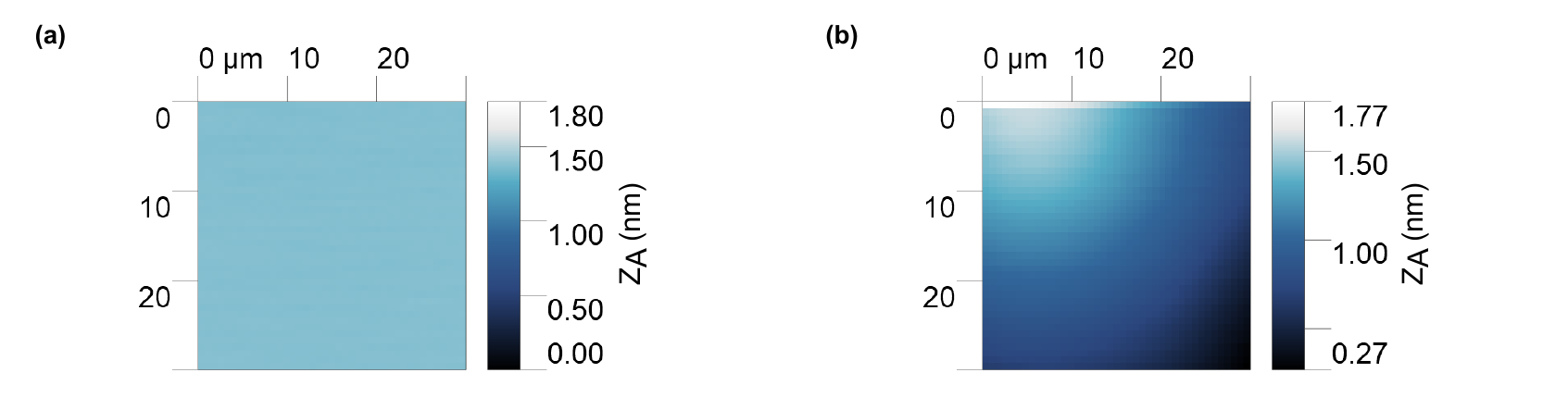} 
    \caption{\textbf{Surface charge imaging.}
    Tip scanned over the surface with (a)~a laser drive and (b)~a tip drive. This data was recorded with a second device at a distance of \SI{1.3}{\micro\meter}.
    }
    \label{fig:SIcharges}
    \end{figure*}



In Fig.~\ref{fig:SIcharges}, we show the amplitude of the antisymmetric mode while scanning over the surface with a constant photon pressure (laser) drive and with a constant tip voltage drive. For the laser drive, we observe a constant amplitude, while the electrically driven amplitude varies significantly. This result is in agreement with our assumption that inhomogeneous surface charges (`voltage patches') are responsible for the electrical interaction of the membrane resonator with the scanning tip. Further systematic studies will hopefully allow us to quantify the charge distribution and to identify potential microscopic origins of the phenomenon. We speculate that such inhomogeneous charge distributions could contribute towards the linear and nonlinear damping of the resonators, as well as to frequency fluctuations~\cite{Moser_2014}.

\section{Second device data}

    \begin{figure*}
    \includegraphics[width=0.5\textwidth]{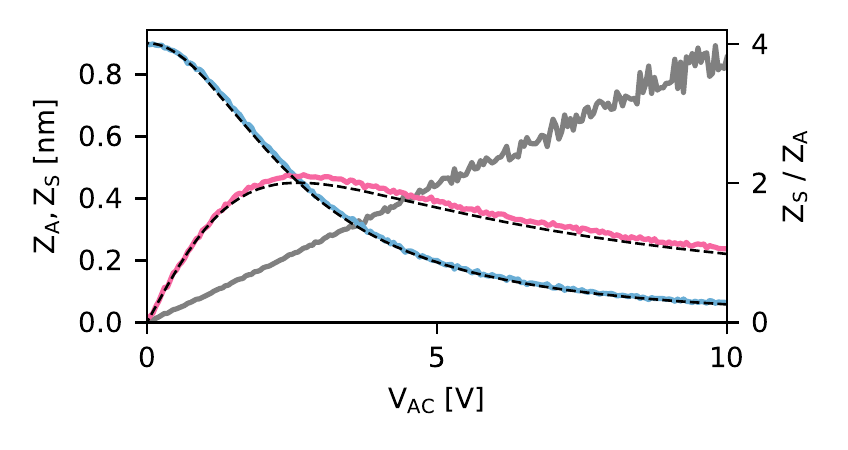} 
    \caption{\textbf{Steady-state amplitudes.}
    Mode amplitudes as a function of $V_\mathrm{AC}$ with a constant photon pressure drive at $f_d = f_A$ and parametric pumping at $f_\Delta$ at a distance $d=$\SI{100}{\nano\meter} (left axis). Blue and pink: measured amplitude of the anti-symmetric and symmetric mode, respectively. Black dashed lines: solutions of Eq.~(3) for $\dot{u}_{S,A} = \dot{v}_{S,A} = 0$. Grey: $Z_S/Z_A$ extracted from the measurements (right axis). This data was recorded with a second device.
    }
    \label{fig:SIsecondDevice}
    \end{figure*}

We repeated the experiment shown in Fig.~4 with a second device with $Q_A=\SI{33e6}{}$ and $Q_S=\SI{31e6}{}$. The steady-state amplitudes were recorded at the position $(0/0)$ of Fig.~\ref{fig:SIcharges}(b) and at a distance of \SI{100}{\nano\meter}. At this position, we find $\frac{g}{2\omega_0} = \SI{0.245}{\radian\per\second}$ for $V_\mathrm{AC} = \SI{10}{\volt}$, which is roughly five times smaller than the coupling measured for the device in the main text under similar conditions.

\section{Slow-flow derivation}

We can model our system as two parametrically coupled, linear oscillators
\begin{align}\label{eq:EOM}
    \ddot{z}_{A} + \Gamma_{A}\dot{z}_{A} + \omega_{A}^2 z_{A} - J z_{S} = F_{A}/m_{A} \\
    \ddot{z}_{S} + \Gamma_{S}\dot{z}_{S} + \omega_{S}^2 z_{S} - J z_{A} = F_{S}/m_{S}\,.
\end{align}
with driving frequency $\Omega$, coupling $g$, dissipation $\Gamma$, displacements $z_A,\, z_S$, angular rotation frequencies $\omega_1,\, \omega_2$, forces $F_A, \, F_S$, and masses $m_A, \, m_S$. We can set the force terms 0 for the moment, and insert them later if needed.

We can split these equations of motion into two first order equations with $z_i = q_j$, $\dot{z}_i=p_j$, $i \in \{A,S\}$ and $j \in \{1,2\}$,
\begin{gather}
    \dot{q}_1 = p_1\label{eq:ODE1_1}\,,\\
    \dot{p}_1 = -\omega_1^2 q_1 + g \cos(\Omega t) q_2 - \Gamma p_1 \label{eq:ODE1_2}\,,\\
    \dot{q}_2 = p_2\label{eq:ODE1_3}\,,\\
    \dot{p}_2 = -\omega_2^2 q_2 + g \cos(\Omega t) q_1 - \Gamma p_2 \label{eq:ODE1_4}\,.  
\end{gather}
We change here from $\{A,S\}$ to $\{1,2\}$ for nicer looking equations, but the indices have nothing to do with the defect site labels 1 and 2. We then move to a rotating frame at an angular frequency $\Omega_1$ for Eqs.~\eqref{eq:ODE1_1} and \eqref{eq:ODE1_2} and at $\Omega_2$ for Eqs.~\eqref{eq:ODE1_3} and \eqref{eq:ODE1_4} using van der Pol transformations~\cite{Papariello_2016}. By integrating the transformed equations over one period, i.e. over $2\pi/\Omega_1$ and $2\pi/\Omega_2$ respectively, we arrive at the equation 

\begin{align}
    \mqty(\dot{u}_1 \\ \dot{v}_1 \\ \dot{u}_2 \\ \dot{v}_2) = \textbf{P} \mqty(u_1 \\v_1 \\u_2 \\v_2)
    = \mqty(
    -\frac{\Gamma }{2} & a_{12} & a_{13} & a_{14}\\
    a_{21} & -\frac{\Gamma }{2} & a_{23} & a_{24}\\
    a_{31} & a_{32} & -\frac{\Gamma }{2} & a_{34}\\
    a_{41} & a_{42} & a_{43} & -\frac{\Gamma }{2}
    )
    \mqty(u_1 \\v_1 \\u_2 \\v_2)
    \label{eq:Propagator}
\end{align}

for the slow-flow coordinates $u_i$ and $v_i$, i.e., the in-phase and out-of-phase oscillation amplitudes. The matrix elements of $\textbf{P}$ are

\begin{align}
    a_{12}=&-a_{21}= \frac{ \Omega_1}{2 \pi } \left(\pi -\frac{\pi   \omega_1^2}{ \Omega_1^2}\right)\, ,\\
    a_{34}=&-a_{43}= \frac{ \Omega_2}{2 \pi } \left(\pi -\frac{\pi   \omega_2^2}{ \Omega_2^2}\right)\, ,\\
    a_{13}=&g \Omega_1 \left(\frac{2 \left(\Omega ^2-\Omega_1^2+\Omega_2^2\right)-(\Omega -\Omega_1+\Omega_2) (\Omega +\Omega_1+\Omega_2) \cos \left(\frac{2 \pi  (\Omega -\Omega_2)}{\Omega_1}\right)}{4 \pi  (\Omega -\Omega_1-\Omega_2) (\Omega +\Omega_1-\Omega_2) (\Omega -\Omega_1+\Omega_2) (\Omega +\Omega_1+\Omega_2)}\right. \nonumber\\
    &\left. -\frac{(\Omega -\Omega_1-\Omega_2) \cos \left(\frac{2 \pi  (\Omega +\Omega_2)}{\Omega_1}\right)}{4 \pi  (\Omega -\Omega_1-\Omega_2) (\Omega -\Omega_1+\Omega_2) (\Omega +\Omega_1+\Omega_2)}\right)\, ,\\
    a_{14}=&\frac{g \Omega_1 \left(\frac{\sin \left(\frac{2 \pi  (\Omega +\Omega_2)}{\Omega_1}\right)}{(\Omega -\Omega_1+\Omega_2) (\Omega +\Omega_1+\Omega_2)}-\frac{\sin \left(\frac{2 \pi  (\Omega -\Omega_2)}{\Omega_1}\right)}{\Omega ^2-2 \Omega  \Omega_2-\Omega_1^2+\Omega_2^2}\right)}{4 \pi }\, ,\\
    a_{23}=&\frac{g \left(\frac{(\Omega -\Omega_2) \sin \left(\frac{2 \pi  (\Omega -\Omega_2)}{\Omega_1}\right)}{(\Omega +\Omega_1-\Omega_2) (-\Omega +\Omega_1+\Omega_2)}-\frac{(\Omega +\Omega_2) \sin \left(\frac{2 \pi  (\Omega +\Omega_2)}{\Omega_1}\right)}{(\Omega -\Omega_1+\Omega_2) (\Omega +\Omega_1+\Omega_2)}\right)}{4 \pi }\, ,\\
    a_{24}=&g\left(\frac{-2 \Omega_2 \left(\Omega ^2+\Omega_1^2-\Omega_2^2\right)+(\Omega -\Omega_2) (\Omega -\Omega_1+\Omega_2) (\Omega +\Omega_1+\Omega_2) \cos \left(\frac{2 \pi  (\Omega -\Omega_2)}{\Omega_1}\right)}{4 \pi  (-\Omega +\Omega_1-\Omega_2) (\Omega +\Omega_1-\Omega_2) (-\Omega +\Omega_1+\Omega_2) (\Omega +\Omega_1+\Omega_2)}\right. \nonumber\\
    &\left. -\frac{(\Omega +\Omega_2) (\Omega -\Omega_1-\Omega_2) \cos \left(\frac{2 \pi  (\Omega +\Omega_2)}{\Omega_1}\right)}{4 \pi  (-\Omega +\Omega_1-\Omega_2) (-\Omega +\Omega_1+\Omega_2) (\Omega +\Omega_1+\Omega_2)}\right)\, ,\\
    a_{31}=& g \Omega_2 \left(\frac{2 \left(\Omega ^2+\Omega_1^2-\Omega_2^2\right)-(\Omega +\Omega_1-\Omega_2) (\Omega +\Omega_1+\Omega_2) \cos \left(\frac{2 \pi  (\Omega -\Omega_1)}{\Omega_2}\right)}{4 \pi  (\Omega -\Omega_1-\Omega_2) (\Omega +\Omega_1-\Omega_2) (\Omega -\Omega_1+\Omega_2) (\Omega +\Omega_1+\Omega_2)}\right. \nonumber\\
    &\left. -\frac{(\Omega -\Omega_1-\Omega_2) \cos \left(\frac{2 \pi  (\Omega +\Omega_1)}{\Omega_2}\right)}{4 \pi  (\Omega -\Omega_1-\Omega_2) (\Omega +\Omega_1-\Omega_2)  (\Omega +\Omega_1+\Omega_2)}\right) \, ,\\
    a_{32}=&\frac{g \Omega_2 \left(\frac{\sin \left(\frac{2 \pi  (\Omega +\Omega_1)}{\Omega_2}\right)}{(\Omega +\Omega_1-\Omega_2) (\Omega +\Omega_1+\Omega_2)}-\frac{\sin \left(\frac{2 \pi  (\Omega -\Omega_1)}{\Omega_2}\right)}{(\Omega -\Omega_1)^2-\Omega_2^2}\right)}{4 \pi }\, ,\\
\end{align}

\begin{align}
    a_{41}=&\frac{g \left(\frac{(\Omega_1-\Omega ) \sin \left(\frac{2 \pi  (\Omega -\Omega_1)}{\Omega_2}\right)}{(\Omega -\Omega_1)^2-\Omega_2^2}-\frac{(\Omega +\Omega_1) \sin \left(\frac{2 \pi  (\Omega +\Omega_1)}{\Omega_2}\right)}{(\Omega +\Omega_1-\Omega_2) (\Omega +\Omega_1+\Omega_2)}\right)}{4 \pi }\, ,\\
    a_{42}=&-g \left(\frac{-2 \Omega_1 \left(\Omega ^2-\Omega_1^2+\Omega_2^2\right)+(\Omega -\Omega_1) (\Omega +\Omega_1-\Omega_2) (\Omega +\Omega_1+\Omega_2) \cos \left(\frac{2 \pi  (\Omega -\Omega_1)}{\Omega_2}\right)}{4 \pi  (\Omega +\Omega_1-\Omega_2) (\Omega -\Omega_1+\Omega_2) (-\Omega +\Omega_1+\Omega_2) (\Omega +\Omega_1+\Omega_2)}\right.\nonumber\\
    &\left. -\frac{(\Omega +\Omega_1) (\Omega -\Omega_1-\Omega_2) \cos \left(\frac{2 \pi  (\Omega +\Omega_1)}{\Omega_2}\right)}{4 \pi  (\Omega +\Omega_1-\Omega_2) (-\Omega +\Omega_1+\Omega_2) (\Omega +\Omega_1+\Omega_2)}\right).
\end{align}

Looking at the off-diagonal $2\times 2$ blocks in $P$ that are responsible for mode coupling, we see that the coupling coefficients are time dependent for most choices of $\Omega_1$ and $\Omega_2$. In our frame rotating at the resonance frequencies of the two modes of interest, a coupling coefficient that varies over time averages out over long enough observation periods and does not contribute significantly. A time-independent, strong coupling can be achieved for $\Omega=\pm(\Omega_1\pm\Omega_2)$. We will therefore only consider the cases of driving at the frequency difference and at the frequency sum.

\subsection{parametric coupling at $\Omega = \Omega_1-\Omega_2$}

For $\Omega = \Omega_1-\Omega_2$ the propagator takes the form 

\begin{align}
    \left(
\begin{array}{cccc}
 -\frac{\Gamma }{2} & \frac{\Omega_1^2-\omega_1^2}{2 \Omega_1} & -\frac{g \Omega_1 \sin ^2\left(\frac{2 \pi  \Omega_2}{\Omega_1}\right)}{8 \pi  \Omega_2 (\Omega_1-\Omega_2)} & \frac{g}{4 \Omega_1}-\frac{g \Omega_1 \sin \left(\frac{4 \pi  \Omega_2}{\Omega_1}\right)}{16 \pi  \Omega_2 (\Omega_1-\Omega_2)} \\
 \frac{(\omega_1-\Omega_1) (\omega_1+\Omega_1)}{2 \Omega_1} & -\frac{\Gamma }{2} & \frac{1}{16} g \left(-\frac{(\Omega_1-2 \Omega_2) \sin \left(\frac{4 \pi  \Omega_2}{\Omega_1}\right)}{\pi  \Omega_2 (\Omega_1-\Omega_2)}-\frac{4}{\Omega_1}\right) & \frac{g (\Omega_1-2 \Omega_2) \sin ^2\left(\frac{2 \pi  \Omega_2}{\Omega_1}\right)}{8 \pi  \Omega_2 (\Omega_1-\Omega_2)} \\
 \frac{g \Omega_2 \sin ^2\left(\frac{2 \pi  \Omega_1}{\Omega_2}\right)}{8 \pi  \Omega_1 (\Omega_1-\Omega_2)} & \frac{g \left(\frac{\Omega_2^2 \sin \left(\frac{4 \pi  \Omega_1}{\Omega_2}\right)}{\pi  \Omega_1^2-\pi  \Omega_1 \Omega_2}+4\right)}{16 \Omega_2} & -\frac{\Gamma }{2} & \frac{\Omega_2^2-\omega_2^2}{2 \Omega_2} \\
 \frac{1}{16} g \left(\frac{(\Omega_2-2 \Omega_1) \sin \left(\frac{4 \pi  \Omega_1}{\Omega_2}\right)}{\pi  \Omega_1 (\Omega_1-\Omega_2)}-\frac{4}{\Omega_2}\right) & \frac{g (2 \Omega_1-\Omega_2) \sin ^2\left(\frac{2 \pi  \Omega_1}{\Omega_2}\right)}{8 \pi  \Omega_1 (\Omega_1-\Omega_2)} & \frac{(\omega_2-\Omega_2) (\omega_2+\Omega_2)}{2 \Omega_2} & -\frac{\Gamma }{2} \\
\end{array}
\right).
\end{align}

The trigonometric terms with $\Omega_1/\Omega_2$ and $\Omega_2/\Omega_1$ lead to micro-motion, but we assume the rotating pictures to be sufficiently close to one another i.e., $\Omega_1/\Omega_2=\Omega_2/\Omega_1=1$. With this approximation, we arrive at

\begin{align}
    \textbf{P} = \left(
\begin{array}{cccc}
 -\frac{\Gamma }{2} & \frac{\Omega_1}{2}-\frac{\omega_1^2}{2 \Omega_1} & 0 & \frac{g}{4 \Omega_1} \\
 \frac{\omega_1^2}{2 \Omega_1}-\frac{\Omega_1}{2} & -\frac{\Gamma }{2} & -\frac{g}{4 \Omega_1} & 0 \\
 0 & \frac{g}{4 \Omega_2} & -\frac{\Gamma }{2} & \frac{\Omega_2}{2}-\frac{\omega_2^2}{2 \Omega_2} \\
 -\frac{g}{4 \Omega_2} & 0 & \frac{\omega_2^2}{2 \Omega_2}-\frac{\Omega_2}{2} & -\frac{\Gamma }{2} \\
\end{array}
\right).
\end{align}

To simplify the notation, we define $\delta_1=\Omega_1-\omega_1$, $\delta_2=\Omega_2-\omega_2$ and $\omega_0=\frac{1}{2}(\omega_1+\omega_2)$. For small detunings, we can set $\frac{\Omega_1}{\omega_1}\approx 1$ and $\frac{\Omega_2}{\omega_2}\approx 1$ and reach the result of equation (2) of the main manuscript

\begin{align}
    \textbf{P} =\left(
\begin{array}{cccc}
 -\frac{\Gamma }{2} & \delta_1 \omega_0 & 0 & \frac{g}{4 \omega_0} \\
 -\delta_1 \omega_0 & -\frac{\Gamma }{2} & -\frac{g}{4 \omega_0} & 0 \\
 0 & \frac{g}{4 \omega_0} & -\frac{\Gamma }{2} & \delta_2 \omega_0 \\
 -\frac{g}{4 \omega_0} & 0 & -\delta_2 \omega_0 & -\frac{\Gamma }{2} \\
\end{array}
\right).
\end{align}

The only difference is that we labeled the symmetric (antisymmetric) mode with $S$ ($A$) instead of $1$ ($2$).

\subsection{parametric coupling at $\Omega = \Omega_1+\Omega_2$}

At $\Omega = \Omega_1+\Omega_2$ the propagator takes the form 

\begin{align}
    \left(
\begin{array}{cccc}
 -\frac{\Gamma }{2} & \frac{\Omega_1^2-\omega_1^2}{2 \Omega_1} & \frac{g \Omega_1 \sin ^2\left(\frac{2 \pi  \Omega_2}{\Omega_1}\right)}{8 \pi  \Omega_2 (\Omega_1+\Omega_2)} & \frac{g \left(\frac{\Omega_1^2 \sin \left(\frac{4 \pi  \Omega_2}{\Omega_1}\right)}{\pi  \Omega_1 \Omega_2+\pi  \Omega_2^2}-4\right)}{16 \Omega_1} \\
 \frac{(\omega_1-\Omega_1) (\omega_1+\Omega_1)}{2 \Omega_1} & -\frac{\Gamma }{2} & \frac{1}{16} g \left(-\frac{(\Omega_1+2 \Omega_2) \sin \left(\frac{4 \pi  \Omega_2}{\Omega_1}\right)}{\pi  \Omega_2 (\Omega_1+\Omega_2)}-\frac{4}{\Omega_1}\right) & \frac{g (\Omega_1+2 \Omega_2) \sin ^2\left(\frac{2 \pi  \Omega_2}{\Omega_1}\right)}{8 \pi  \Omega_2 (\Omega_1+\Omega_2)} \\
 \frac{g \Omega_2 \sin ^2\left(\frac{2 \pi  \Omega_1}{\Omega_2}\right)}{8 \pi  \Omega_1 (\Omega_1+\Omega_2)} & \frac{g \left(\frac{\Omega_2^2 \sin \left(\frac{4 \pi  \Omega_1}{\Omega_2}\right)}{\pi  \Omega_1^2+\pi  \Omega_1 \Omega_2}-4\right)}{16 \Omega_2} & -\frac{\Gamma }{2} & \frac{\Omega_2^2-\omega_2^2}{2 \Omega_2} \\
 \frac{1}{16} g \left(-\frac{(2 \Omega_1+\Omega_2) \sin \left(\frac{4 \pi  \Omega_1}{\Omega_2}\right)}{\pi  \Omega_1 (\Omega_1+\Omega_2)}-\frac{4}{\Omega_2}\right) & \frac{g (2 \Omega_1+\Omega_2) \sin ^2\left(\frac{2 \pi  \Omega_1}{\Omega_2}\right)}{8 \pi  \Omega_1 (\Omega_1+\Omega_2)} & \frac{(\omega_2-\Omega_2) (\omega_2+\Omega_2)}{2 \Omega_2} & -\frac{\Gamma }{2} \\
\end{array}
\right)
\end{align}

Using a similar set of assumption as in the previous case, we obtain

\begin{align}
    \textbf{P} =\left(
\begin{array}{cccc}
 -\frac{\Gamma }{2} & \delta_1 \omega_0 & 0 & -\frac{g}{4 \omega_0} \\
 -\delta_1 \omega_0 & -\frac{\Gamma }{2} & -\frac{g}{4 \omega_0} & 0 \\
 0 & -\frac{g}{4 \omega_0} & -\frac{\Gamma }{2} & \delta_2 \omega_0 \\
 -\frac{g}{4 \omega_0} & 0 & -\delta_2 \omega_0 & -\frac{\Gamma }{2}\,. \\
\end{array}
\right)
\end{align}

\section{Derivation of Eq.~(4) and (5)}
For the experiment presented in Fig.~4, the phases of both modes are depend on the external force. Equations~(4) and (5) can be derived as follows: an external force $F_A$ drives the symmetric mode to the standard linear response
\begin{align}
Z_A
=
\frac{F_A}{m\omega_0\Gamma}e^{-\frac{\pi}{2}}
=
-i \frac{F_A}{m\omega_0\Gamma}\,.
\end{align}
Adding a coupling term $J z_A$ with the component $\frac{1}{2}g Z_A \cos(\omega_S t)$ acting resonantly on $Z_S$, we obtain via the same mechanism
\begin{align}
Z_S
=
\frac{g Z_A}{2\omega_0\Gamma}e^{-\frac{\pi}{2}}
=
-i \frac{g Z_A}{2\omega_0\Gamma}\,,
\end{align}
which we divide by $Z_A$ to obtain Eq.~(5). $Z_S$ exerts a coupling force $g Z_S/2$ back on $z_A$. The steady-state equation we find for $Z_A$ is
\begin{align}
Z_A = -i \frac{F_A}{m\omega_0\Gamma} -i \frac{g Z_S}{2\omega_0\Gamma} = -i \frac{F_A}{m\omega_0\Gamma} - \left(\frac{g}{2\omega_0\Gamma}\right)^2 Z_A \,,
\end{align}
which we solve for $Z_A$ to obtain Eq.~(4).

\section{Driving at the frequency sum}


    \begin{figure}
    \includegraphics[width=0.5\columnwidth]{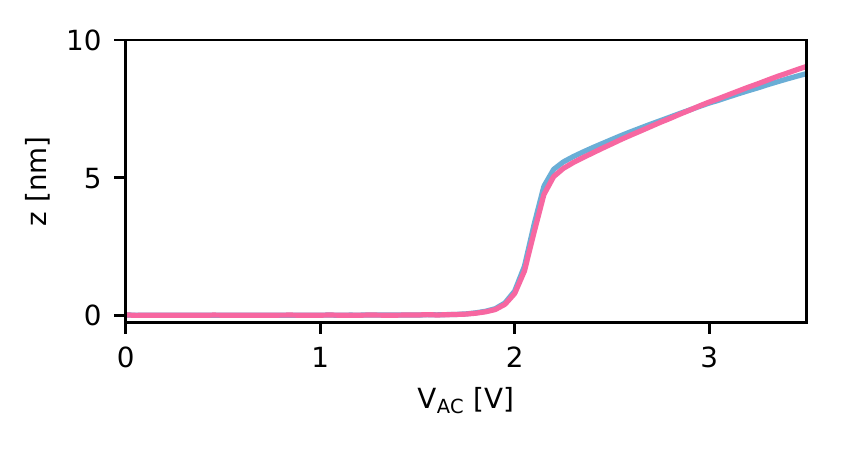} 
    \caption{\textbf{Driving at the frequency sum.} Experimental result of driving the two modes at the sum of their frequencies $f_A+f_S$ with a tip-voltage drive and with $d = \SI{1000}{\nano\meter}$.
    }
    \label{fig:Sumdrive}
    \end{figure}

We also tested parametric driving at the frequency sum $f_\Sigma=f_A+f_S$. Here, owing to the symmetry in the propagation matrix in Eq.~(3), the two modes drive each other simultaneously instead of alternatingly. As a result, there is a parametric driving threshold $g_\mathrm{th}$ where the energy pumped into the system exceeds the energy lost through linear damping~\cite{Olkhovets_2001,Zhang_2002}. This threshold is reached when the gain factor
\begin{align}\label{eq:G}
	G = \left[\left(1 + \frac{m g}{2}\sqrt{\frac{Q_S Q_A}{k_S k_A}}\right)\left(1 - \frac{m g}{2}\sqrt{\frac{Q_S Q_A}{k_S k_A}}\right)\right]^{-1}
\end{align}
diverges~\cite{Olkhovets_2001}. With $m = m_A \approx m_S$, $Q_A \approx Q_S$ and $k_A = m\omega_A^2 \approx k_S = m\omega_S^2$, this is the case for $\frac{g_\mathrm{th}}{2\omega_0} = \Gamma$. With the calibrations from Fig.~3 and 4, we expect this threshold around $V_{AC} = \SI{2.5}{\volt}$.

In Fig.~\ref{fig:Sumdrive} we observe the amplitude response of the two modes to a parametric sum drive in the absence of an external force. The amplitudes, measured with a narrow-band filter in our lock-in amplifier, remain zero below the threshold at roughly $V_{AC} = \SI{1.8}{\volt}$. Beyond the threshold, both amplitudes rise quickly and reach a high amplitude limited by the nonlinearity, similar to a parametric oscillator~\cite{Lifshitz_2008}. The difference between the expected threshold ($V_{AC} = \SI{2.5}{\volt}$) and the measured result could be due to slow drift of the nanopositioner relative to measurements shown in the main text.

\providecommand{\noopsort}[1]{}\providecommand{\singleletter}[1]{#1}%